\documentclass[12pt]{article}

\newcommand{\mrm}[1]{\mathrm{#1}}
 
\newcommand{\gap}{\stackrel{>}{\sim}}

\newcommand{\alphas}{\alpha_{\mrm{s}}}
\newcommand{\alphaem}{\alpha_{\mrm{em}}}

\newcommand{\pT}{p_{\perp}}
\newcommand{\kT}{k_{\perp}}

\renewcommand{\a}{\mrm{a}}
\renewcommand{\b}{\mrm{b}}
\renewcommand{\c}{\mrm{c}}
\renewcommand{\d}{\mrm{d}}
\newcommand{\e}{\mrm{e}}

\newcommand{\g}{\mrm{g}}

\newcommand{\p}{\mrm{p}}
\newcommand{\q}{\mrm{q}}
\newcommand{\s}{\mrm{s}}

\renewcommand{\u}{\mrm{u}}

\newcommand{\J}{\mrm{J}}

\newcommand{\bbar}{\overline{\mrm{b}}}
\newcommand{\cbar}{\overline{\mrm{c}}}
\newcommand{\dbar}{\overline{\mrm{d}}}

\newcommand{\qbar}{\overline{\mrm{q}}}
\newcommand{\sbar}{\overline{\mrm{s}}}

\newcommand{\ubar}{\overline{\mrm{u}}}

\newcommand{\qqbar}{\q\qbar}
\newcommand{\uubar}{\u\ubar}
\newcommand{\ddbar}{\d\dbar}
\newcommand{\ssbar}{\s\sbar}

\newenvironment{Itemize}{\begin{list}{$\bullet$}%
{\setlength{\topsep}{0.2mm}\setlength{\partopsep}{0.2mm}%
\setlength{\itemsep}{0.2mm}\setlength{\parsep}{0.2mm}}}%
{\end{list}}
\newcounter{enumct}
\newenvironment{Enumerate}{\begin{list}{\arabic{enumct}.}%
{\usecounter{enumct}\setlength{\topsep}{0.2mm}%
\setlength{\partopsep}{0.2mm}\setlength{\itemsep}{0.2mm}%
\setlength{\parsep}{0.2mm}}}{\end{list}}
 
\newlength{\captivewidth}
\newlength{\abstwidth}
\def\thebibliographys#1{\subsection*{References}\list
  {[\arabic{enumi}]}{\settowidth\labelwidth{[#1]}\leftmargin\labelwidth
    \advance\leftmargin\labelsep
    \usecounter{enumi}}
    \def\newblock{\hskip .11em plus .33em minus -.07em}
    \sloppy
    \sfcode`\.=1000\relax}

\begin{document}
 
\sloppy

\pagestyle{empty}

\begin{flushright}
CERN--TH/95--62  \\
LU TP 95--6 \\
March 1995
\end{flushright}
 
\vspace{\fill}
 
\begin{center}
{\LARGE\bf Low- and high-mass components of}\\[4mm]
{\LARGE\bf the photon distribution functions}\\[10mm]
{\Large Gerhard A. Schuler$^a$} \\[3mm]
{\it Theory Division, CERN} \\[1mm]
{\it CH-1211 Geneva 23, Switzerland}\\[1mm]
{ E-mail: schulerg@cernvm.cern.ch}\\[2ex]
{\large and} \\[2ex]
{\Large Torbj\"orn Sj\"ostrand} \\[3mm]
{\it Department of Theoretical Physics  2}\\[1mm]
{\it University of Lund, Lund, Sweden}\\[1mm]
{ E-mail: torbjorn@thep.lu.se}
\end{center}
 
\vspace{\fill}
 
\begin{center}
{\bf Abstract}\\[2ex]
\begin{minipage}{\abstwidth}
The structure of the general solution of the inhomogeneous evolution 
equations allows the separation of a photon structure function  
into perturbative (``anomalous") and non-perturbative contributions. 
The former part is fully calculable, and can be identified with the 
high-mass contributions to the dispersion integral in the photon mass. 
Properly normalized ``state" distributions can be defined, where 
the $\gamma\to\qqbar$ splitting probability is factored out. 
These state distributions are shown to be useful in the description 
of the hadronic event properties, and necessary for a proper 
eikonalization of jet cross sections. Convenient parametrizations
are provided both for the state and for the full anomalous parton
distributions. The non-perturbative parts of the parton distribution 
functions of the photon are identified with the low-mass contributions 
to the dispersion integral. Their normalizations, as well as the value 
of the scale $Q_0$ at which the perturbative parts vanish, are fixed by 
approximating the low-mass contributions by a discrete, finite sum of 
vector mesons. The shapes of these hadronic distributions are fitted to 
the available data on $F_2^\gamma(x,Q^2)$. Parametrizations are provided 
for $Q_0=0.6\,$GeV and $Q_0=2\,$GeV, both in the DIS and the 
$\overline{\mathrm{MS}}$ factorization schemes. The full parametrizations
are
extended towards virtual photons. Finally, the often-used 
``FKP-plus-TPC/$2\gamma$" solution for $F_2^\gamma(x,Q^2)$ is 
commented upon. 
\end{minipage}
\end{center}

\vspace{\fill}
\noindent
\rule{6cm}{0.4mm}

\vspace{1mm} \noindent
${}^a$ Heisenberg Fellow.
\hfill\\[1ex]


\noindent
CERN--TH/95--62

\clearpage
\pagestyle{plain}
\setcounter{page}{1}

\section{Introduction}

Inclusive cross sections, such as 
deep-inelastic electron--photon scattering 
$\e \gamma \rightarrow \e + X$ or high-$\pT$ jet production in 
two-photon collisions
$\gamma\gamma \rightarrow \mathrm{jet} + X$, are fully expressed 
in terms of the parton distribution functions (PDFs) of the photon 
$f_a^{\gamma}(x,Q^2)$ (for recent reviews on the photon structure 
functions see e.g.\ refs.\ 
\cite{Berger87,Kolanoski88,Vieira91,Borzumati93,Vogt94,Brodsky94}).  
Correspondingly, such measurements allow, 
in principle, a unique determination of the PDFs of the photon. 
However, already for the simplest determination, namely the extraction 
of $F_2^{\gamma}(x,Q^2)$ from deep-inelastic electron--photon 
scattering, additional information about the hadronic final state is 
necessary \cite{Miller94}. 
The reason is simple: in contrast to lepton--nucleon scattering,
where $(x,Q^2)$ are fixed by the measurement of the outgoing lepton 
solely, in $e\gamma$ scattering $x$ usually can only be determined 
through a reconstruction of the hadronic energy $W$ from the 
visible one by virtue of the relation $x = Q^2/(W^2 + Q^2)$. 
Clearly, an exclusive description of the event 
properties is not only of interest for acceptance corrections but 
provides also additional insight in the structure of the photon. 
 
In this paper we shall investigate the photon-to-$\qqbar$ splitting 
probability and define (normalized) ``state" distributions 
where this probability factor is split off. We shall show that the use 
of these state distributions is of great advantage 
in obtaining a correct description 
of the hadronic final state, notably of the 
photon remnants and the initial-state gluon radiation 
in parton-shower programs. We shall see that the state distributions  
enter also the calculation of jet cross sections at high energies 
where unitarity corrections have to be taken into account. 

Furthermore, we shall show how to obtain a constrained 
parametrization of the (inclusive) PDFs of the photon. 
Current parametrizations either are fully unconstrained, i.e.\ 
all parameters are fitted to $F_2^\gamma$ data \cite{LAC}, or 
rely on severe prejudices
about the input distributions, such as approximating these  
by the ones of the pion \cite{GRV,AFG}
or modelling them by the quark--parton results  
(box-diagram) with the quark masses as free parameters \cite{GS}. 
By contrast, the shapes of our input distributions are determined by the 
$F_2^\gamma$ data, while the normalizations are constrained. 
There exists a one-to-one relation between the size of the non-perturbative 
part and the scale $Q_0$ introduced to separate the non-perturbative 
and the perturbative parts. This scale should be universal, 
i.e.\ process-independent and can, for example, be constrained by
a study of the $\gamma\p$ total cross section \cite{Schuler93}. 
We also investigate the sensitivity of the PDFs to the 
photonic factorization scheme associated with the presence of a 
direct contribution $C^\gamma$ to $F_2^\gamma$. 
The current analysis updates our old parametrization of the 
photon PDFs \cite{Schuler93}. 

Next we discuss the dependence of the PDFs of the photon on its mass
(virtuality) $P^2$, i.e.\ the so-called target-mass effects.
Our parametrization can, in fact, also be used for the 
PDFs of a virtual photon, and we shall outline the assumptions 
that are behind this. Effects of a non-zero virtuality have 
attracted considerable interest 
\cite{Walsh,Uematsu,Borzumati93,Kim93,Drees94,Aurenche94,Gluck94}, 
but no explicit (analytic) parametrization was available up to now. 

The outline of the paper is as follows.
In the next section we first discuss the photon-to-$\qqbar$ 
splitting and the issues of introducing a scale $Q_0$ 
to separate low-mass contributions from perturbatively 
calculable high-mass contributions. The latter, the so-called 
anomalous contributions, are then expressed in terms of ``state 
distributions" that enter the description of the hadronic 
event (section~2.1), the eikonalization of jet cross sections 
(section~2.2), and the virtual PDFs (section~2.3). 
Section~3 contains details about the parametrizations of the 
parameter-free state distributions and the anomalous distributions. 
The small-$x$ behaviour of the distribution functions is outlined. 
In section~4 we discuss the normalization of the 
non-perturbative (low-mass) contributions to the 
photonic PDFs (``hadronic distributions") 
and compare our fits to the $F_2^\gamma$ data. 
Finally we comment upon the often-used approximation of $F_2^\gamma$ 
by the sum of the ($Q^2$-independent) TPC/$2\gamma$ parametrization 
\cite{TPC} and the QCD expression of FKP \cite{FKP}.
Our results are summarized in section~5. 
There we also discuss further aspects of our separation of the 
photon distribution functions into long-distance and short-distance 
components, in particular the connection with the representation 
of a photonic PDF as a dispersion integral in the photon mass. 

\section{Separation of the non-perturbative part}

Perturbative QCD predicts only the $Q^2$ evolution of the 
PDFs of the photon $f_i^{\gamma}(x,Q^2)$, via a set of inhomogeneous 
differential equations of the first kind. Hence the solutions 
$f_i^{\gamma}(x,Q^2)$ require the specification of the PDFs 
at some $Q^2 = Q_0^2$. It is well known that these 
``input" distributions $f_i^{\gamma}(x,Q_0^2)$ have to have 
a considerable ``hard" component if $Q_0$ is chosen to be a 
typical hadronic input scale ($Q_0 \gap 2\,$GeV). This is quite 
in contrast to hadronic PDFs which vanish in a powerlike way $\propto 
(1-x)^p$ ($p\gap 1$) for $x \rightarrow 1$. Correspondingly, there 
is no unique guiding principle as to what the photonic input distributions 
should look like. Rather, parametrizations starting at 
large $Q_0$ \cite{LAC,GS}
are obtained by just fitting the parameters of the 
input distributions 
$f_a^{\gamma}(x,Q_0^2)$ (shapes and normalizations)
to the available $F_2^{\gamma}$ data.   
Moreover, since these data are so far restricted to large $x$, 
basically no constraint on the gluon distribution of the photon 
exists today in such an approach. 

Lowering the starting scale $Q_0$ to about $0.5$--$0.7\,$GeV,  
it is, however, possible to describe the $F_2^{\gamma}$ data 
with photonic PDFs obtained from a purely hadronic input  
\cite{GRV,Schuler93,AFG}. The latter is usually estimated 
by the PDFs of the pion using vector-meson-dominance (VMD) 
and the additive quark model. 
The recent next-to-leading-order 
parametrizations AFG \cite{AFG} and GRV \cite{GRV}  
essentially differ in the ansatz to take the photon 
to be a coherent \cite{AFG} or incoherent \cite{GRV} 
superposition of vector mesons. 
Both are parametrizations of the inclusive PDFs of the photon. 
Our aim \cite{Schuler93} is to obtain PDFs 
where the probability for the photon 
to fluctuate into a hadronic state is made explicit. 
Firstly, this allows us to constrain the normalizations 
of the non-perturbative input distributions, 
while the shapes are fitted to $F_2^\gamma$ data. 
Secondly, we can trace the complete evolution of the 
parton showering, otherwise hidden in the expression 
of $f_i^{\gamma}(x,Q^2)$. Thirdly, differences in the 
remnant structure of the photon can be respected. Finally, 
our parametrizations hold also for a virtual photon, to which
they can easily be applied because the parametrizations  
are analytic in both $P^2$ and $Q^2$. 
On the other hand, our parametrizations are leading-order ones only.  

Analogously to our decomposition of the $\gamma\p$ total cross section 
\begin{equation}
 \sigma_{\mrm{tot}}^{\gamma\p} = 
   \sigma_{\mrm{dir}}^{\gamma\p} + 
   \sigma_{\mrm{VMD}}^{\gamma\p} + 
   \sigma_{\mrm{anom}}^{\gamma\p}  
\ , 
\label{sigmagamp}
\end{equation}
we decompose the PDFs of the photon as
\begin{equation}
 f_a^{\gamma}(x,Q^2) = f_a^{\gamma,\mrm{dir}}(x,Q^2)
  + f_a^{\gamma,\mrm{VMD}}(x,Q^2,Q_0^2) 
  + f_a^{\gamma,\mrm{anom}}(x,Q^2;Q_0^2) 
\ .
\label{gammaPDF}
\end{equation}
The ansatz (\ref{sigmagamp}), or eq.~(\ref{gammaPDF}), is based on 
the assumption that there exists a scale $Q_0$ which divides the 
spectrum of $\gamma\to \qqbar$ fluctuations: 
above $Q_0$ these can be described 
perturbatively\footnote{A similar definition of the 
  perturbative part of $F_2^\gamma(x,Q^2)$ is due to FKP \cite{FKP}.  
  A comparison is postponed to section~\ref{FKPsec}.}
(``anomalous" term in eqs.~(\ref{sigmagamp}) and (\ref{gammaPDF})), 
while below $Q_0$ the fluctuations are assumed to give vector-meson
states. 

A reminder of the familiar scale ambiguity problem may be in place.
In this paper we work throughout with the deep-inelastic-scattering
convention in mind, where the $Q^2$ scale is set by the virtuality
of the probing photon. Fluctuations $\gamma \leftrightarrow \q\qbar$
of the probed photon are resolved as long as the ``scale'' $k$ of
these fluctuations is below $Q$. However, the precise physics
definition of $k$ is left unspecified. When PDFs extracted from 
deep inelastic scattering are applied to other processes, such as 
$\q\q' \to \q\q'$ high-$\pT$ jet production, the choice of relevant
scale $Q$ is ambiguous. It could be associated with the transverse
momentum $\pT$ of the jets, or some multiple thereof. Conversely, our
transverse momentum cut-off, used in \cite{Schuler93} to separate
between direct, simple VMD (i.e.\ $\rho^0$, $\omega$, and $\phi$)
and anomalous processes, is of the order of the
$Q_0 \sim 0.6$~GeV used in this paper, but the two could well differ
by some amount. 

The first term in eq.~(\ref{gammaPDF}) describes the 
(properly normalized) probability 
distribution of a photon to remain a photon
\begin{equation}
f_a^{\gamma,\mrm{dir}}(x,Q^2) = Z_3 \, \delta_{a\gamma} \, 
\delta (1-x)
\ ,
\label{dirgammaPDF}
\end{equation}
where 
\begin{equation}
 Z_3 = 1 - \sum_{a=\q,\qbar,\g}\; \int_0^1\, {\d} x\; x\ 
  \left\{ f_a^{\gamma,\mrm{VMD}}(x,Q^2,Q_0^2) 
  + f_a^{\gamma,\mrm{anom}}(x,Q^2;Q_0^2) \right\}
\ .
\label{Zthreedef}
\end{equation}
Properly speaking, also fluctuations 
$\gamma \leftrightarrow \ell^+\ell^-$, $\ell = \e, \mu$ or $\tau$,
should be taken into account for $Z_3$. This contribution is fully
perturbatively calculable. In practice, $Z_3 \approx 1$ is a 
sufficiently good approximation for all applications.

The PDFs of the photon, being the solution of an inhomogeneous 
evolution equation (or, more precisely, a system of equations) 
can always be written as the sum of two terms 
\begin{equation}
f_a^{\gamma}(x,Q^2) -  f_a^{\gamma,\mrm{dir}}(x,Q^2)
  = f_a^{\gamma,\mrm{PT}}(x,Q^2,Q_0^2) 
   + f_a^{\gamma,\mrm{NP}}(x,Q^2,Q_0^2)
\ .
\label{generaldeco}
\end{equation} 
The first term is a particular solution of the {\em inhomogeneous}  
equation with the boundary condition 
\begin{equation}
f_a^{\gamma,\mrm{PT}}(x,Q_0^2,Q_0^2) = 0
\ .
\label{BCanom}
\end{equation}
The second term,  
a general solution of the corresponding {\em homogeneous} 
evolution equation 
needs a (non-perturbative) input distribution at $Q^2=Q_0^2$
\begin{equation}
f_a^{\gamma,\mrm{NP}}(x,Q_0^2,Q_0^2) =  F_a^{\mrm{NP}}(x)
\ .
\label{BCNP}
\end{equation}

While mathematically the decomposition (\ref{generaldeco}) is valid 
for any $Q_0^2$, for physics reasons we want to identify 
$f_a^{\gamma,\mrm{PT}}$ with the 
perturbatively calculable distributions $f_a^{\gamma,\mrm{anom}}$
arising from the point-like coupling 
of the photon to a quark--antiquark pair. 
These distributions can be expressed as an integral of 
``state" distribution functions $f_a^{\gamma,\q\qbar}(x,Q^2;k^2)$: 
\begin{equation}
f_a^{\gamma,\mrm{anom}}(x,Q^2;Q_0^2) = 
\frac{\alphaem}{2\pi} \, \sum_{\q} 2 e_{\q}^2 \, 
\int_{Q_0^2}^{Q^2} \frac{{\d} k^2}{k^2} \, 
f_a^{\gamma,\q\qbar}(x,Q^2;k^2) ~,
\label{angamsol}
\end{equation}
which obey the standard, homogeneous evolution equations 
with the boundary condition
\begin{equation}
f_a^{\gamma,\q\qbar}(x,k^2;k^2) = f_a^{\gamma,\q\qbar}(x) 
 \equiv \frac{3}{2} \, \left( x^2 + (1-x)^2 \right) \,
( \delta_{a\q} + \delta_{a\qbar} ) 
\ .
\label{fxinit}
\end{equation}

The solution (\ref{angamsol}) 
can be understood as follows. The probability for the
photon to branch into a $\q\qbar$ state at some high 
scale $k$ is given by
$(\alphaem/2\pi) \, 2e_{\q}^2 \, {\d} k^2/k^2$. Once the photon has  
fluctuated, the possibility of additional QCD evolution between $k$
and $Q$ is just given by the standard evolution equations. This
evolution conserves total momentum, so that  
\begin{equation}
\sum_a \, \int_0^1 {\d} x \, x f_a^{\gamma,\q\qbar}(x, Q^2; k^2) = 1
\qquad ({\mrm{for}}~Q \geq k) 
\ ,
\label{momsumevol}
\end{equation}
where the sum now runs over the gluon and all the quark and antiquark 
species. 

Similarly, $f_a^{\gamma,\mrm{NP}}$ should describe the partonic content 
of the photon that has fluctuated to a vector meson, i.e.\ 
%
\begin{equation}
 f_a^{\gamma,\mrm{VMD}}(x,Q^2,Q_0^2) 
    =  \sum_V \frac{4\pi\alphaem}{f_V^2} f_a^{\gamma,V}(x,Q^2,Q_0^2)
 ~.
\label{VMDgammaPDF}
\end{equation}
Here $4\pi\alphaem/f_V^2$ gives the probability for the photon 
to fluctuate into the vector meson $V$, while $f_a^{\gamma,V}$ obeys
a momentum sum rule just as $f_a^{\gamma,\q\qbar}$ in 
eq.~(\ref{momsumevol}). Normally we let the sum run over $\rho^0$,
$\omega$ and $\phi$, with the first responsible for the bulk of the
contribution. 

In passing we note that eq.~(\ref{momsumevol}) yields
\begin{equation}
  Z_3 = 1 -  \sum_V \frac{4\pi\alphaem}{f_V^2} 
   - \frac{\alphaem}{2\pi} \, \sum_{\q} 2 e_{\q}^2 \, 
\ln \left( \frac{Q^2}{Q_0^2} \right)
\ .
\label{Zthreevalue}
\end{equation}
It is convenient to think of the anomalous component as a
continuous spectrum of states,
characterized by the flavour $\q$ and the scale of creation $k$,
just as the VMD components give a discrete spectrum of states.
The probability $(\alphaem/2\pi) \, 2e_{\q}^2 \, {\d} k^2/k^2$ 
in eq.~(\ref{angamsol}) is
therefore the equivalent of the VMD couplings $4\pi\alphaem/f_V^2$
in eq.~(\ref{VMDgammaPDF}).   

\subsection{Hadronic event properties}

For inclusive quantities, such as the high-$\pT$ jet rate in 
$\gamma\gamma$ collisions, only the full parton
distributions $f_a^{\gamma}(x,Q^2)$ are of interest. 
However, for an exclusive description of event properties, 
states with different initial flavours and $k$ values behave 
differently \cite{PWZ83,Schuler93}. 
A physically transparent description of the complete hadronic
final state therefore requires that $f_a^{\gamma}(x,Q^2)$ be
available subdivided into the $f_a^{\gamma,\mrm{anom}}(x,Q^2)$
and $f_a^{\gamma,\mrm{VMD}}(x,Q^2)$ parts, with these in 
their turn subdivided as an integral or sum of state distributions
$f_a^{\gamma,\q\qbar}(x,Q^2)$ and $f_a^{\gamma,V}(x,Q^2)$.
Let us elaborate on this.

1. Hard processes (at a scale $\pT \sim Q$) in 
photon-induced reactions are given by 
the usual $2\to 2$ partonic processes, where the parton $a$ in the 
photon is selected according to the inclusive distributions 
$f_a^{\gamma}(x,Q^2)$. In addition there are the processes where
the photon interacts directly, 
i.e.\ via $f_a^{\gamma,\mrm{dir}}(x,Q^2)$. 
For consistency, the direct cross sections have to be cut off at 
the same $Q_0$ as is used to define the anomalous PDFs, i.e.\ 
the cut-off will, in general, be different for different parametrizations 
and should be known!

2. The $2\to 2$ partonic processes are then supplemented by initial 
(and final) state parton showers. In the case of partons in the photon, 
the inhomogeneous evolution equations have to be used. However, 
parametrizations of the inclusive PDFs of the photon are not 
guaranteed to be positive-definite at the infrared cutoff 
used in the parton-shower programmes. Nor is the shape at $Q_0$ 
constrained in any way to correspond to a hadronic distribution, so that 
it makes sense to approximate the photon remnants by those of a hadron. 
On the other hand, both demands are fulfilled if
$f_a^{\gamma}(x,Q^2)$ is split  
as in eqs.~(\ref{gammaPDF}), (\ref{angamsol}) and (\ref{VMDgammaPDF}). 
Moreover, existing parton shower programmes for homogeneous evolution, 
made and tested for hadron-induced reactions, can then be used, since 
both $f_a^{\gamma,\q\qbar}(x,Q^2;k^2)$ and $f_a^{\gamma,V}(x,Q^2)$ 
evolve according to the homogeneous equations. 
The separation (\ref{gammaPDF}) also ensures that the 
``intrinsic" transverse momentum $\kT$ of the initial parton 
in the photon can be correctly generated: 
a roughly Gaussian primordial $k_{\perp}$ distribution of
width $\sim 0.5\,$GeV for low-mass hadronic fluctuations of the photon,
and a ${\d} k_{\perp}^2/k_{\perp}^2$ distribution for high-mass 
fluctuations. 

\subsection{Eikonalization of jet cross-sections}

The state distributions 
$f_a^{\gamma,\q\qbar}(x,Q^2;k^2)$ and $f_a^{\gamma,V}(x,Q^2)$ 
enter also the calculation of total photon-induced cross sections 
at high energies. There (mini)jet cross sections rise much faster 
than allowed by the Froissart bound and unitarization corrections 
have to be taken into account. Unitarization effects can be estimated 
by the eikonal formula. Respecting the probability
for the photon to be in different hadronic fluctuations, 
the correct formula for e.g.\ $\gamma\p$ interactions is
\begin{eqnarray}
\sigma^{\gamma\p}_ {\mrm{inel}}(s) & = & \sigma_{\mrm{dir}}^{\gamma\p}(s)
+  \sum_V \frac{4\pi\alphaem}{f_V^2}\; \int {\d}^2 b
 \left\{1 - \exp\left[ -2\, \mathrm{Im} \chi^{V\p}(s,b) \right] \right\}
\nonumber\\ & & \quad
 + \frac{\alphaem}{2\pi} \sum_{\q} 2e_{\q}^2 \int_{Q_0^2}
 \frac{{\d} k_{\perp}^2}{k_{\perp}^2} \;  \int {\d}^2 b
 \left\{1 - \exp\left[ -2\, \mathrm{Im} \chi^{\qqbar\p}(s,b,k_{\perp}^2)
\right] \right\} 
\ .
\label{gammaeikonal}
\end{eqnarray} 
The state distributions enter the hard cross sections, e.g.\ 
\begin{equation}
\sigma_{\mathrm{hard}}^{\qqbar\p}(s;\kT) =
\int {\d} x_1 \int {\d} x_2 \int {\d} \pT^2
  f_{i}^{\gamma,\q\qbar}(x_1,\pT^2;\kT) f_{j}^{\p}(x_2,\pT^2)
  \frac{{\d}\hat{\sigma}_{ij}} {{\d} \pT^2}
  \theta( \pT - \kT)
\ ,
\label{minijet}
\end{equation}
which in turn enter the eikonals $\chi^i$ 
in eq.~(\ref{gammaeikonal}), e.g.\ 
$\chi^{\qqbar\p} = A^{\qqbar\p} \sigma^{\qqbar\p}$, where
$A^{\qqbar\p}(b,k_{\perp}^2)$ is an overlap
function and $\sigma^{\qqbar\p}(s,k_{\perp}^2) =
\sigma^{\qqbar\p}_{\mathrm{soft}}(s,k_{\perp}^2) + 
\sigma^{\qqbar\p}_{\mathrm{hard}}(s,k_{\perp}^2)$.

\subsection{PDFs of a virtual photon}

Finally, the separation (\ref{gammaPDF}) is useful 
to study the effects of a non-zero photon virtuality $P^2$. 
Generalized vector-meson dominance suggests the following dispersion 
relation in $k^2$:  
\begin{equation}
  f_a^{\gamma^\star}(x,Q^2,P^2) = \int_0^{Q^2} \frac{{\d} k^2}{k^2}\; 
  \left( \frac{k^2}{k^2 + P^2} \right)^2 \; 
    \frac{\alphaem}{2\pi} \, \sum_{\q} 2 e_{\q}^2 \, 
        f_a^{\gamma,\q\qbar}(x,Q^2;k^2) 
\ .
\label{virtualGVD}
\end{equation}
The integration from zero to $Q_0^2$ can be approximated 
by the (low-mass) vector-meson contributions, which show the 
usual fast $P^2$ fall-off predicted by pole dominance: 
\begin{equation}
 f_a^{\gamma^\star,\mrm{VMD}}(x,Q^2,P^2) = 
   \sum_V\; \left(\frac{m_V^2}{m_V^2 + P^2}\right)^2 
  \; \frac{4\pi\alphaem}{f_V^2}\;  f_a^{\gamma,V}(x,Q^2,P_0^2)
\ .
\label{virtualVMD}
\end{equation}
The pole-dominance factor 
takes the reduced probability for a virtual photon to fluctuate 
into a vector meson into account. On the other hand, the PDFs 
of a virtual vector meson are completely unknown. 
The simplest choice, namely to identify them with those of a real 
pion,  i.e.\ taking $P_0^2 = Q_0^2$ in eq.~(\ref{virtualVMD}) 
\cite{Gluck94}, 
leads to unphysically large hadronic contributions at small $x$. 
Moreover, this choice does not guarantee the vanishing of the PDFs 
as $P^2 \rightarrow Q^2$, as should be the case up to power-suppressed
terms. Our choice $P_0^2 = \max(P^2,Q_0^2)$ 
is based both on this limit and on the expectation that the PDFs 
should become more valence-like as $P^2$ increases. 

The anomalous contribution is given by the integration $k^2 > Q_0^2$ 
in eq.~(\ref{virtualGVD}) and we approximate it by 
\begin{eqnarray}
 f_a^{\gamma^\star,\mrm{anom}}(x,Q^2,P^2) & = &  
  \int_{\max{(P^2,Q_0^2)}}^{Q^2} \frac{{\d} k^2}{k^2}\; 
    \frac{\alphaem}{2\pi} \, \sum_{\q} 2 e_{\q}^2 \, 
        f_a^{\gamma,\q\qbar}(x,Q^2;k^2) 
\nonumber\\
  & \equiv &  f_a^{\gamma,\mrm{anom}}(x,Q^2,P_0^2) 
\ .
\label{virtualanom}
\end{eqnarray}
The ansatz (\ref{virtualanom}) guarantees that the virtual PDFs  
\begin{equation}
 f_a^{\gamma^\star}(x,Q^2,P^2) = f_a^{\gamma,\mrm{dir}}(x,Q^2,P^2)
  + f_a^{\gamma^\star,\mrm{VMD}}(x,Q^2,P^2)
  + f_a^{\gamma^\star,\mrm{anom}}(x,Q^2;P^2) 
\label{virtualPDF}
\end{equation} 
are correct in the region $\Lambda^2 \ll P^2 \ll Q^2$: 
there the virtual PDFs can be calculated exactly
and QCD  predicts them to be given by eq.~(\ref{angamsol}) 
with $Q_0^2$ replaced by $P^2$ \cite{Uematsu}.  
The PDFs of eq.~(\ref{virtualPDF}) are exact also in the limit 
$P^2 \rightarrow 0$ where they approach the real PDFs. 
 
On the other hand, for $P^2$ of the order of $Q_0^2$,
i.e. in the interesting ``cross-over'' region $P^2 \sim m_{\rho}^2$, 
the $P^2$ dependence 
of the anomalous contribution is less certain. A power-like 
dependence might be present that could be estimated via
eq.~(\ref{virtualGVD}). Our choice of $P_0^2$ in 
eq.~(\ref{virtualanom}) 
is motivated by (i) simplicity and (ii) the demand for a smooth transition 
towards higher $P^2$, but also alternatives like $P_0^2 = P^2 + Q_0^2$
would have fulfilled these demands. Uncertainties arising from this 
choice may be estimated by varying $P_0^2$, say, from $\max(P^2,Q_0^2/2)$ 
to $\max(P^2,2 Q_0^2)$.

As $P^2$ approaches $Q^2$, the concept of virtual PDFs breaks down 
since powerlike corrections $\propto (P^2/Q^2)^p$ then become 
important besides the logarithmic ones. Nevertheless, the 
logarithmically enhanced terms of the $P^2 \rightarrow Q^2$ 
limit exhibit the correct behaviour: the quark distributions 
approach the box-diagram expression 
(quark--parton model result) \cite{Walsh},
and the gluon distribution vanishes faster than the quark ones 
\cite{Borzumati93}. 
We finally want to emphasize that our parametrization 
of the virtual PDFs is analytic in all variables and can hence 
be used easily. 
It is well known that hadronic distribution functions, such 
as $f_a^{\gamma,V}(x,Q^2,P_0^2)$, depend on the two momentum scales 
$Q^2$ and $P_0^2$ only through the logarithmic integration of 
the coupling constant (see eq.~(\ref{sdef}) below). However, also 
the solution of the inhomogeneous equation can be written in a similar 
form, see eq.~(\ref{fulldist}) below. 

\section{The perturbative part}

\subsection{The state distributions}

The distributions $f_a^{\gamma,\q\qbar}(x,Q^2;k^2)$ depend on two
momentum scales, $Q$ and $k$. However, the amount of evolution that 
occurs between these two scales is entirely characterized by the 
logarithmic integration of the strong coupling constant 
\begin{equation} 
s = s(Q^2,k^2) = \int_{k^2}^{Q^2} \frac{{\d} Q^2}{Q^2} \,
\frac{\alphas(Q^2)}{2\pi} 
~.
\label{sdef}
\end{equation}
In leading order
$s= (1/b) \, \ln[ \ln(Q^2/\Lambda^2) / \ln(k^2/\Lambda^2) ]$, 
where $b=(33-2 n_f)/6$. 
We have here assumed that the number of flavours is fixed,
e.g. $n_f = 4$, but it is straightforward to split the $Q^2$ range 
into subranges with different $n_f$ values and correspondingly
matched $\Lambda^{(n_f)}$ values. The number of flavours also enters 
in the fraction of the momentum carried by gluons to that carried by 
quarks; here a matching between different $n_f$ values is less 
transparent, but can still be attempted. 

The parametrization of the quark distributions is divided into a
valence and a sea part. The sea part corresponds to the distributions
$f_a^{\gamma,\q\qbar}$ for $a\neq\q,\qbar$, while the valence
part is the additional contribution obtained for $a=\q$ or $a=\qbar$,
i.e.
\begin{eqnarray}
f_{\q}^{\gamma,\q\qbar} = f_{\qbar}^{\gamma,\q\qbar} & = &
f_{\q,\mrm{val}}^{\gamma,\q\qbar} + f_{\q,\mrm{sea}}^{\gamma,\q\qbar}
\nonumber\\
f_{\q'}^{\gamma,\q\qbar} = f_{\qbar'}^{\gamma,\q\qbar} & = &
f_{\q,\mrm{sea}}^{\gamma,\q\qbar} \qquad \mrm{for}~ \q' \neq \q  
\ . 
\label{separ}
\end{eqnarray}

As always in QCD, the solutions are too complex to be given in closed
form. One therefore has to resort to approximate parametrizations.
We perform an evolution in $x$ space 
choosing a strategy similar to the one proposed by Odorico \cite{Odorico},
wherein a large number of evolution histories are traced by Monte
Carlo methods. The evolved parton 
densities are binned in $x$ for several $Q^2$ values. Thereafter
the $x$ shape is parametrized in some simple form, and the $Q^2$
dependence of these parameters is in turn parametrized.
The {\sc Minuit} program \cite{James} is used to find suitable 
parametrization coefficients, but with frequent manual interaction 
to ensure a sensible behaviour. 
The fits have been made in the range $10^{-4} \leq x \leq 1$ and 
$Q_0 \leq Q \leq 2000 Q_0$ (where $Q_0 \approx 0.5\,$GeV), 
but the forms have been chosen 
so that they can be used for all $x$. In regions where the distributions
are large, the typical accuracy of the parametrizations is 1\%--2\%. 

The parametrizations are
\begin{eqnarray}
x f_{\q,\mrm{val}}^{\gamma,\q\qbar}(x,s) & = &
\left( c_1^{\mrm{val}} \, x^2 + c_2^{\mrm{val}} \, 
(1-x)^2 + c_3^{\mrm{val}} \, x(1-x) \right) 
x^{\displaystyle c_4^{\mrm{val}}} \, 
(1-x^2)^{\displaystyle c_5^{\mrm{val}}}
\nonumber\\
x f_{a,\mrm{sea}}^{\gamma,\q\qbar}(x,s) & = & c_1^{\mrm{sea}} \, 
x^{\displaystyle c_2^{\mrm{sea}}} \, 
(1-x)^{\displaystyle c_3^{\mrm{sea}}} \,
x f_{a,\mrm{sea},\mrm{lo}}^{\gamma,\q\qbar}(x)
\nonumber\\
x f_{\g}^{\gamma,\q\qbar}(x,s) & = & c_1^{\mrm{g}} \, 
x^{\displaystyle c_2^{\mrm{g}}} \, 
((1-x)(- \ln x))^{\displaystyle c_3^{\mrm{g}}} \,
x f_{\g,\mrm{lo}}^{\gamma,\q\qbar}(x) 
\ , 
\label{stateparam}
\end{eqnarray}
where
\begin{eqnarray}
x f_{a,\mrm{sea},\mrm{lo}}^{\gamma,\q\qbar}(x) & = &
\frac{8 - 73 x + 62 x^2}{9} (1-x) + 
\left( \frac{8 x^2}{3} - 3 \right) x \ln x + (2x - 1) x \ln^2 x 
\label{seadist}
\\
x f_{\g,\mrm{lo}}^{\gamma,\q\qbar}(x) & = &
\frac{4 + 7 x + 4 x^2}{3} (1-x) + 2x(1+x) \ln x 
\label{gludist} 
\end{eqnarray}
and the coefficients $c_i$ are given in table~\ref{statetable}. 
\begin{table}
\begin{center}
\begin{tabular}{|c|r|r|r|r|r|r|r|}
\cline{2-8} 
\multicolumn{1}{c|}{} & $a$ & $b$ & $c$ & $d$ & $e$ & $f$ & $g$ 
\\ \hline
$c_1^{\mathrm{val}}$ & $1.5$ & -- & -- & $-0.197$ & $4.33$ & -- & -- 
\\ \hline
$c_2^{\mathrm{val}}$ & $1.5$ & $2.10$ & -- & $3.29$ & -- & -- & -- 
\\ \hline
$c_3^{\mathrm{val}}$ & -- & $5.23$ & -- & $1.17$ & -- & $19.9$ & -- 
\\ \hline
$c_4^{\mathrm{val}}$ & $1$ & -- & -- & $1.5$ & -- & -- & -- 
\\ \hline
$c_5^{\mathrm{val}}$ & -- & $2.667$ & -- & -- & -- & -- & -- 
\\ \hline
$c_1^{\mathrm{sea}}$ & -- & -- & $1$ & $4.54$ & $8.19$ & $8.05$ & -- 
\\ \hline
$c_2^{\mathrm{sea}}$ & -- & $1.54$ & -- & $1.29$ & -- & -- & -- 
\\ \hline
$c_3^{\mathrm{sea}}$ & -- & $2.667$ & -- & -- & -- & -- & -- 
\\ \hline
$c_1^{\mathrm{g}}$ & -- & $4$ & -- & $4.76$ & $15.2$ & -- & $29.3$ 
\\ \hline
$c_2^{\mathrm{g}}$ & -- & $2.03$ & -- & $2.44$ & -- & -- & -- 
\\ \hline
$c_3^{\mathrm{g}}$ & -- & $1.333$ & -- & -- & -- & -- & -- 
\\ \hline
\end{tabular}
\caption{Coefficients of the state distributions parametrized in the 
form \protect\newline
$c_i = (a+bs + cs^2)/(1+ds+es^2+fs^3+gs^4)$.}
\end{center}
\label{statetable}
\end{table}

Note that the shapes and normalizations of the parametrizations 
become exact in the limit $s \rightarrow 0$: the 
expression in eq.~(\ref{fxinit}) is recovered in the
limit $s \to 0$ of the valence distribution, and the gluon 
(sea) distribution approaches 
eq.~(\ref{gludist}) (eq.~(\ref{seadist})) times 
$4s$ ($s^2$). Hence, the sea distribution vanishes faster than the 
gluon distribution, which in turn vanishes faster than the 
valence one \cite{Borzumati93}.  
This is because $4sf_{\g,\mrm{lo}}^{\gamma,\q\qbar}(x)$ is 
obtained if the initial
quark-distribution in eq.~(\ref{fxinit}) is convoluted with the 
$\q \to \q\g$ splitting kernel. Similarly,  
$s^2 f_{a,\mrm{sea},\mrm{lo}}^{\gamma,\q\qbar}(x)$
is obtained if the first-order gluon-distribution
in eq.~(\ref{gludist}) is convoluted with the $\g \to \q\qbar$ 
splitting kernel.  
Moreover, our parametrizations exhibit the correct large-$x$ behaviour, 
$(1-x)^{8s/3}$ for the valence distribution, 
$(1-x)^{1+8s/3}$ for the gluon one, and 
$(1-x)^{2+8s/3}$ for the sea one. 

Analytic results can also be derived for the limit $x \rightarrow 0$. 
We find
\begin{eqnarray}
x f_{\q,\mrm{val}}^{\gamma,\q\qbar}(x,s) & \rightarrow &
  \frac{3}{2}\; \exp{\left(\frac{2s}{3}\right)}\; 
  I_0(z_v)
\nonumber\\
x f_{\g}^{\gamma,\q\qbar}(x,s) & \rightarrow &
  \frac{8}{9}\; h(s)\, \sqrt{\frac{6s}{y}}\; I_1(z_g)
\nonumber\\
x f_{a,\mrm{sea}}^{\gamma,\q\qbar}(x,s) & \rightarrow &
  \frac{8}{27}\; h(s)\, \frac{s}{y}\; I_2(z_s)
\ ,
\label{smallx}
\end{eqnarray} 
where $h(s) = \exp{[(-297-2n_f)s/54]}$, $z_v=2\sqrt{4sy/3}$, 
$z_s = z_g = 2\sqrt{6ys}$, and $y=\ln(1/x)$. 
Since $I_{\nu}(z) \rightarrow \exp(z)/\sqrt{2\pi z}$ for 
$z\rightarrow \infty$, the gluon and sea distributions grow 
as $\exp{2\sqrt{6s\ln(1/x)}}$ at small $x$. 
However, for realistic $x$ and $Q^2$ values, 
the distributions are far from being asymptotic, and the power-like
parametrizations (\ref{stateparam}) are better suited. 
  
The effects of quark masses have not been taken into account in
the evolution. For charm (and bottom), the sea distributions should 
therefore be modified. Different levels of sophistication can be
used. The simplest is to assume that branchings $\g \to \c\cbar$
are forbidden at scales below the charm mass, while $m_{\c}$ can be 
neglected in the evolution above threshold. In this context, $m_{\c}$
can be thought of as an effective parameter, to be adjusted 
for taking into account threshold effects \cite{Yuri}. 
In the region of small $s$, the number of gluons increases like $s$, 
that of light sea quarks like $s^2$, and that of charm sea quarks like 
$s^2 - s_{\c}^2$, with $s_{\c} = s(\max(k^2,m_{\c}^2),k^2)$. For 
simplicity, one may thus assume
\begin{equation}
f_{\c,\mrm{sea}}^{\gamma,\q\qbar}(x,s) = \theta(s-s_{\c}) \,
\left( 1 - \left( \frac{s_{\c}}{s} \right)^2 \right) \,
f_{\a,\mrm{sea}}^{\gamma,\q\qbar}(x,s) ~,
\label{charmsupp}
\end{equation}
where $\theta(x)$ is the ordinary step function.

Also original splittings $\gamma \to \c\cbar$ (and 
$\gamma \to \b\bbar$) should be suppressed by mass thresholds.
Again assuming a simple step threshold, one obtains
$f_a^{\gamma,\c\cbar}(x, Q^2; k^2) = 0$ for $Q < m_{\c}$,
while the distribution for $Q > m_{\c}$ is given by
the substitution
\begin{equation}
s = s(Q^2, k^2) \to s(Q^2, \max(k^2, m_{\c}^2)) ~.
\end{equation} 

\subsection{The anomalous distributions}

It would have been very convenient, had it been possible, to generate
the full distributions in a simple way from the state ones above.
Unfortunately, the expressions are too complicated to allow the
${\d} k^2/k^2$ integral in eq.~(\ref{angamsol}) to be carried out.
The parametrization of $f_a^{\gamma,\mrm{anom}}$
is therefore done separately from the ones given above.

The $f_a^{\gamma,\mrm{anom}}(x,Q^2;Q_0^2)$ distributions do depend 
both on the $Q/\Lambda$ and the $Q_0/\Lambda$ ratio, 
and not just on a single variable. 
However, it is possible to separate the two dependences 
and to obtain a parametrization that is analytic in both 
$Q$ and $Q_0$. Hence, our parametrization can easily be 
used for a virtual photon target. 

To this end we introduce the functions 
$\overline{f}_a^{\gamma,\q\qbar}$ defined by: 
\begin{equation}
\overline{f}_a^{\gamma,\q\qbar}(x,s_0) = \int_0^{s_0}\, b\, {\d} s \, 
\frac{\exp{\left(- b s \right)}}{1 - \exp{\left(-b s_0\right)} }
 \, f_a^{\gamma,\q\qbar}(x,s)
\ .
\label{fbardef}
\end{equation}
Here $s_0$ is defined analogously to $s$ in eq.~(\ref{sdef}), i.e.\ 
$s_0= (1/b) \, \ln[ \ln(Q^2/\Lambda^2) / \ln(Q_0^2/\Lambda^2) ]$. 
The $\overline{f}_a^{\gamma,\q\qbar}$ distributions can be seen as a 
weighted mean of the $f_a^{\gamma,\q\qbar}$ ones; in particular
$\overline{f}_a^{\gamma,\q\qbar}$ has a unit 
momentum sum, 
\begin{equation}
\sum_a \, \int_0^1 {\d} x \, x \overline{f}_a^{\gamma,\q\qbar}(x,s_0) = 
1 ~. 
\label{momsumfull}
\end{equation}
The information on the fraction of photons that are split by 
$\gamma \to \q\qbar$ is thus only carried by the prefactor
proportional to $\ln(Q^2/Q_0^2)$ so that: 
\begin{equation}
x f_a^{\gamma,\mrm{anom}}(x,Q^2;Q_0^2) = 
\frac{\alphaem}{2\pi} \, \sum_{\q} 2 e_{\q}^2 \, 
\ln \left( \frac{Q^2}{Q_0^2} \right) \,
x \overline{f}_a^{\gamma,\q\qbar}(x,s_0) 
~.
\label{fulldist}
\end{equation}  

The $\overline{f}_a^{\gamma,\q\qbar}$ have been parametrized in the 
same spirit as the $f_a^{\gamma,\q\qbar}$ ones, with due respect
to the limiting behaviours for $s \to 0$.  
For the valence and sea parametrizations we take the 
same functional form as in eq.~(\ref{stateparam}), 
but for the gluon  we choose 
\begin{equation}
x \overline{f}_{\g}^{\gamma,\q\qbar}(x,s_0)  =  c_1^{\g} \, 
x^{\displaystyle c_2^{\g}} \, (1-x^2)^{\displaystyle c_3^{\g}} \,
x f_{\g,\mrm{lo}}^{\gamma,\q\qbar}(x)
 ~.
\end{equation}
This form is not exactly like that for the gluon state-distribution, rather 
$((1-x)(- \ln x))^{\displaystyle c_3}$ is replaced by 
$(1-x^2)^{\displaystyle c_3}$.
We have attempted to use the same form in both cases, but then failed
to obtain reasonable fits. The limit $s \to 0$ is again exact.
The coefficients $c_i$ are collected in table~\ref{fulltable}. 
\begin{table}
\begin{center}
\begin{tabular}{|c|r|r|r|r|r|r|r|}
\cline{2-8} 
\multicolumn{1}{c|}{}& $a$ & $b$ & $c$ & $d$ & $e$ & $f$ & $g$ 
\\ \hline
$c_1^{\mathrm{val}}$ & $1.5$ & $2.49$ & $26.9$ & -- & $32.3$ & -- & -- 
\\ \hline
$c_2^{\mathrm{val}}$ & $1.5$ & $-0.49$ & $7.83$ & -- & $7.68$ & -- & -- 
\\ \hline
$c_3^{\mathrm{val}}$ & -- & $1.5$ & -- & $-3.2$ & $7.$ & -- & -- 
\\ \hline
$c_4^{\mathrm{val}}$ & $1$ & -- & -- & $0.58$ & -- & -- & -- 
\\ \hline
$c_5^{\mathrm{val}}$ & -- & $2.5$ & -- & $10.$ & -- & -- & -- 
\\ \hline
$c_1^{\mathrm{sea}}$ & -- & -- & $0.333$ & $4.90$ & $4.69$ & $21.4$ & -- 
\\ \hline
$c_2^{\mathrm{sea}}$ & -- & $-1.18$ & -- & $1.22$ & -- & -- & -- 
\\ \hline
$c_3^{\mathrm{sea}}$ & -- & $1.22$ & -- & -- & -- & -- & -- 
\\ \hline
$c_1^{\mathrm{g}}$ & -- & $2$ & -- & $4.$ & $7.$ & -- & -- 
\\ \hline
$c_2^{\mathrm{g}}$ & -- & $-1.67$ & -- & $2.$ & -- & -- & -- 
\\ \hline
$c_3^{\mathrm{g}}$ & -- & $1.2$ & -- & -- & -- & -- & -- 
\\ \hline
\end{tabular}
\end{center}
\caption{Coefficients of the full distributions parametrized 
as in table~\protect\ref{statetable}.}
\label{fulltable}
\end{table}

The small-$x$ behaviour is again better described by a powerlike 
distribution over the relevant $x$ ranges, even though asymptotically 
the growth is weaker, for example 
\begin{equation}
 x\  f_{\g}^{\gamma,\mrm{anom}}(x,Q^2,Q_0^2) 
  \stackrel{x \rightarrow 0}{\rightarrow} \frac{\alphaem}{2\pi}\; 
  \left( \sum_f 2 e_f^2 \right)\, \ln\frac{Q^2}{\Lambda^2}\; 
  \frac{8 b s}{9 y}\; \exp\left(- \frac{297-8n_f}{27}\, s \right)\; 
  I_2(z_g) 
\ .
\label{gasympt}
\end{equation}

The charm (and bottom) seas are suppressed; close to
threshold by a relative amount $1 - (s_{\c}/s)^3$, where
$s_{\c} = s(\max(Q_0^2,m_{\c}^2),Q_0^2)$. (Note that this
factor approaches unity faster than the factor in 
eq.~(\ref{charmsupp}), as a consequence of having a spectrum 
of initial branchings at different $k^2$ scales.)
For $Q > m_{\c}$, the charm sea is therefore assumed to be given by the
normal sea multiplied by this factor.

From the above formulae, the full distribution is built up by
summing the contributions from the allowed flavours as already 
specified in eq.~(\ref{fulldist}).  
Branchings $\gamma \to \c\cbar$ (and $\gamma \to \b\bbar$) do not
contribute below threshold. Above threshold, the charm contribution
is obtained by substituting $\max(Q_0^2,m_{\c}^2)$ for $Q_0^2$ in 
the logarithmic prefactor and in the definition of $s$. 

\subsection{The heavy-flavour contribution}

The above treatment of heavy-flavour branchings 
$\gamma \to \c\cbar, \b\bbar$ is not unreasonable for jet production
in $\gamma\gamma$ collisions of real photons, $Q^2 = P^2 =0$
(with the r\^ole of hard scale here taken by the transverse mass of 
the quarks).
However, in the deep-inelastic-scattering region, 
$Q^2 \gg P^2 \approx 0$, the kinematics constraint 
$W^2 = Q^2 (1-x) / x > 4 m_{\c}^2$ implies that the distribution can
only be non-vanishing for $x < Q^2/(Q^2 + 4m_{\c}^2)$. The large 
charm-quark charge ensures that the charm contribution is very significant
for the shape of the $F_2^{\gamma}(x, Q^2)$ distribution. 

For the charm contribution to $F_2^{\gamma}$, we have therefore based
ourselves
on the leading-order ``Bethe--Heitler'' cross section
for $\gamma^* \gamma^* \to \c\cbar$. 
(An additional charm contribution comes from our sea parametrization, 
i.e.\ from $\g \to \c\cbar$ branchings.)
The full expression is given
in \cite{Budnev}, but for our applications it is sufficient to use
the approximation in \cite{Hill}, valid in the limit
$4 x^2 P^2 \ll Q^2$:
\begin{eqnarray}
x c(x,Q^2) = \frac{\alphaem}{2\pi} 3 \left( \frac{2}{3} \right)^2 \,
x  & & \left\{ \rule[-2ex]{0mm}{5ex}
  \beta_{\c} \left[ 6x (1-x) -1 \right] \right. \nonumber \\ 
& & + \log \left( \frac{1+\beta_{\c}\eta_{\c}}{1-\beta_{\c}%
\eta_{\c}} \right)  \left[ x^2 + (1-x)^2 + r_{\c} x (1-3x) - 
\frac{1}{2} r_{\c}^2 x^2 \right]  \nonumber \\
& & \left. + \frac{2x}{Q^2} \frac{2\beta_{\c}\eta_{\c}}%
{1-\beta_{c}^2\eta_{\c}^2}
\left[ (2-r_{\c}) m_{\c}^2 x - P^2 x \right] \right\} ~,
\end{eqnarray}
with
\begin{equation}
r_{c} = \frac{4 m_{\c}^2}{Q^2} ~, \quad \quad
\beta_{\c} = \sqrt{ 1 - \frac{4 m_{\c}^2}{W^2} } ~, \quad \quad
\eta_{\c} = \sqrt{ 1 - \frac{4 x^2 P^2}{Q^2} } ~, 
\end{equation}  
and correspondingly for $\b$.

Some distance above the threshold, the Bethe--Heither formula gives
essentially the same contribution to $F_2^{\gamma}$ as does the standard
formulae presented in the previous section.

In the calculations, we have used fairly small $\c$ and $\b$ masses,
$m_{\c} = 1.3$~GeV and $m_{\b} = 4.6$~GeV, so that the charm contribution
is large. On the other hand, the $\J/\psi$ and $\Upsilon$ states are not
included in the VMD sum over vector mesons. To first approximation,
these two effects should therefore cancel. The approach can be motivated 
by standard duality arguments.

\section{The non-perturbative part}

\subsection{Normalization}

We now turn to the hadronic input distributions 
$f_a^{\gamma,V}(x) = f_a^{\gamma,V}(x,Q_0^2,Q_0^2)$, 
cf.\ eq.~(\ref{VMDgammaPDF}).
Usually these are estimated by the pion PDFs using 
VMD and the additive quark model.  
Of course, the PDFs of the $\qqbar$ ``bound states" of the photon 
need not be the same as those of real vector mesons. 
Moreover, the PDFs of the short-lived $\rho^0$-meson 
may well differ in shape from those of the long-lived pion. Therefore, 
in contrast to the GRV \cite{GRV} and AFG \cite{AFG} parametrizations,
we do not approximate the hadronic input distributions $f_a^{\gamma,V}(x)$ 
by the pion input distributions $f_a^{\pi}(x)$, but rather 
constrain the shape of the input distributions from data. 
 
We do, however, use VMD to fix the normalization of the 
input distributions. Indeed, for consistency with eq.~(\ref{sigmagamp}) 
we cannot allow for an extra ``K-factor" as is introduced in 
refs.\ \cite{GRV,AFG} if the input scale $Q_0$ corresponds to 
the simple $\rho^0$, $\omega$, $\phi$ VMD model.
Based on eqs.~(\ref{generaldeco}), (\ref{angamsol}) and
(\ref{VMDgammaPDF}) one has 
\begin{eqnarray}
\lefteqn{f_a^{\gamma}(x,Q^2) -  f_a^{\gamma,\mrm{dir}}(x,Q^2)
  = }
\nonumber\\ & & 
 \sum_V \frac{4\pi\alphaem}{f_V^2} f_a^{\gamma,V}(x,Q^2,Q_0^2)
  + \frac{\alphaem}{2\pi} \, \sum_{\q} 2 e_{\q}^2 \, 
\int_{Q_0^2}^{Q^2} \frac{{\d} k^2}{k^2} \, 
f_a^{\gamma,\q\qbar}(x,Q^2;k^2) 
\ .
\label{VMDsum}
\end{eqnarray} 
The analogous decomposition (\ref{sigmagamp}) 
of the $\gamma$p total cross section 
tells us that $Q_0$ should be of the order of 
$600\,$MeV if the sum over vector mesons includes only the three 
lowest-lying states $V_1 = \rho^0$, $\omega$, and $\phi$. On the other 
hand, for larger values of $Q_0$ more vector mesons $V_n$ 
have to be included: 
\begin{equation}
 \sum_{V=\rho^0,\omega,\phi} \frac{e}{f_V} | V \rangle \Longrightarrow 
   \sum_{n=1}^{N(Q_0)}\, \sum_{V=\rho^0,\omega,\phi} 
    \frac{e}{f_{V_n}} | V_n \rangle 
\ .
\end{equation}
Since the VMD couplings of such higher-mass states are poorly 
known, mass effects become more important. Moreover,
the input distributions will anyhow be fitted, so we may approximate 
\begin{equation}
 \sum_{n=1}^{N(Q_0)}\, \sum_{V=\rho^0,\omega,\phi} 
    \frac{4\pi\alphaem}{f_{V_n}^2} 
   f_a^{\gamma,V_n}(x) \approx  K\, 
     \sum_{V=\rho^0,\omega,\phi} 
    \frac{4\pi\alphaem}{f_{V}^2}  
   \tilde{f}_a^{\gamma,V}(x)
\ .
\label{Kfactor}
\end{equation}
In principle, for a given $Q_0$,  
$K = K(Q_0)$ could be determined by repeating the $\gamma p$ total 
cross-section analysis. In the following we 
will consider two cases, (i) $Q_0=0.6\,$GeV and $K=1$ fixed from the 
$\gamma p$ analysis, and (ii) $Q_0 = 2\,$GeV, where $K$ will be fitted 
together with the input distributions $\tilde{f}(x)$ to the 
$F_2^\gamma(x,Q^2)$ data.

The normalization of the PDFs discussed so far 
still leaves us freedom of how to add the vector mesons. For example, 
in the case of simple VMD: 
\begin{equation}
|\gamma\rangle_{\mrm{VMD}} = \sum_{V=\rho,\omega,\phi}\, 
  \frac{e}{f_V}\,  | V \rangle = 
 \sqrt{\frac{e^2}{f_{\rho}^2} + \frac{e^2}{f_{\omega}^2}}\; 
  \left(e_{\u}^2 + e_{\d}^2\right)^{-1/2}\; 
  \left( e_{\u}\, |\uubar\rangle 
      + e_{\d}\, |\ddbar\rangle \right) 
 + \frac{e}{f_{\phi}}\,  | \ssbar \rangle
\ .
\label{PDFdecompo}
\end{equation} 
The coherent superposition of vector mesons 
is obtained for $e_{\u} = 2/3$, $e_{\d} = -1/3$.  
A completely $SU_3$-symmetric coherent superposition would give
$f_{\rho}/f_{\omega} = 1/3$ and 
\mbox{$f_{\rho}/f_{\phi} = - \sqrt{2}/3$}.
This is close to the experimental numbers, but in particular $\phi$
production seems suppressed compared to the above expectation.
This is not unreasonable in view of the larger $\s$ quark and
$\phi$ meson masses. We therefore keep the VMD couplings at
their measured values, which we also use 
for the total cross section in eq.~(\ref{sigmagamp}).
An incoherent superposition of $\u$ and $\d$ quarks corresponds to  
$e_{\u} = e_{\d} = 1$. At long time scales, e.g. in ``elastic''
processes such as $\gamma\p \to V \p$, the $\rho^0$ and $\omega$ vector
mesons contain equal amounts of $\u\ubar$ and $\d\dbar$, i.e. the
coherence at the $\gamma\q\qbar$ vertex is broken. It is therefore 
conceivable to have either physics scenario, or some intermediate
thereof, and both have been used in the literature  
\cite{Berger87,GRV,AFG}. Here we favour the coherent superposition of 
$\u$ and $\d$ quarks, in line with the argument that hard processes 
probe short time scales.

\begin{table}
\begin{center}
\begin{tabular}{|l|cl|cl|cl|}
\hline
$a$ 
 & \multicolumn{2}{c|}{Incoherent} 
 & \multicolumn{2}{c|}{Coherent}  
 & \multicolumn{2}{c|}{Coherent (naive $SU_3$)} 
\\ \hline
u-val &  
  $(g_{\rho}^2 + g_{\omega}^2)/2$    & $\approx 0.248$ & 
  $4 (g_{\rho}^2 + g_{\omega}^2)/5$  & $\approx 0.397$ & 
  $8 g_{\rho}^2/9$                   & $\approx 0.404$
\\ \hline 
u-sea & 
  $g_{\rho}^2 + g_{\omega}^2 + g_{\phi}^2$  & $\approx 0.551$ & 
  $g_{\rho}^2 + g_{\omega}^2 + g_{\phi}^2$  & $\approx 0.551$ & 
  $4g_{\rho}^2/3$                    & $\approx 0.606$
\\ \hline
d-val &  
  $(g_{\rho}^2 + g_{\omega}^2)/2$    & $\approx 0.248$ & 
  $(g_{\rho}^2 + g_{\omega}^2)/5$  & $\approx 0.099$ & 
  $2 g_{\rho}^2/9$                   & $\approx 0.101$
\\ \hline 
d-sea & 
  $g_{\rho}^2 + g_{\omega}^2 + g_{\phi}^2$  & $\approx 0.551$ & 
  $g_{\rho}^2 + g_{\omega}^2 + g_{\phi}^2$  & $\approx 0.551$ & 
  $4g_{\rho}^2/3$                    & $\approx 0.606$
\\ \hline
s-val &  
  $g_{\phi}^2$  & $\approx 0.054$ & 
  $g_{\phi}^2$  & $\approx 0.054$ & 
  $2 g_{\rho}^2/9$                   & $\approx 0.101$
\\ \hline 
s-sea & 
  $g_{\rho}^2 + g_{\omega}^2 + g_{\phi}^2$  & $\approx 0.551$ & 
  $g_{\rho}^2 + g_{\omega}^2 + g_{\phi}^2$  & $\approx 0.551$ & 
  $4g_{\rho}^2/3$                    & $\approx 0.606$
\\ \hline 
gluon & 
  $g_{\rho}^2 + g_{\omega}^2 + g_{\phi}^2$  & $\approx 0.551$ & 
  $g_{\rho}^2 + g_{\omega}^2 + g_{\phi}^2$  & $\approx 0.551$ &  
  $4g_{\rho}^2/3$                    & $\approx 0.606$
\\ \hline
\end{tabular}
\caption{The coefficients $c\protect_a$ 
of (\protect\ref{tabulated}) for coherent 
and incoherent superpositions of vector mesons.
Here $g_V^2 = 4\pi/f_V^2$.
\label{coeff}}
\end{center}
\end{table}
For the input distributions we assume 
an $SU_3$-symmetric sea distribution $s(x)$ 
and denote by $v(x)$ the input valence distribution 
such that $\Sigma \equiv \sum_{\q} (x q(x) + x \bar{q}(x) ) 
= 2 v(x) + 6 s(x)$. The coefficients 
of the various hadronic input distributions $f_a^{\gamma,\mrm{VMD}}(x)$ 
are summarized in table~\ref{coeff} for the three possibilities 
of photon decomposition discussed above, and for $K=1$. 
We have used an obvious notation, e.g.\ 
\begin{equation}
 f_{a,\mrm{val}}^{\gamma,\mrm{VMD}}(x,Q_0^2,Q_0^2) = 
\alphaem\, c_{a,\mrm{val}}\,v(x)
\ ,
\label{tabulated}
\end{equation} 
and our preferred choice corresponds to the middle column
in table~\ref{coeff}. 
For completeness, we list also the expression of $F_2$ in terms 
of $s(x)$ and $v(x)$ for this choice 
\begin{eqnarray}
  \frac{1}{\alphaem}\; F_2^{\gamma,\mrm{VMD}}(x,Q^2) & = & 
 \left[ \frac{34}{45}\, \left( \frac{4\pi}{f_{\rho}^2} +  
    \frac{4\pi}{f_{\omega}^2} \right) 
       + \frac{2}{9}\, \frac{4\pi}{f_{\phi}^2}
   \right]\; x\ v(x) + 
  \frac{4}{3}\, \sum_{V=\rho,\omega,\phi} 
   \frac{4\pi}{f_{V}^2} \; x\ s(x)
\nonumber\\
  & = & 0.3875\, x\ v(x) + 0.735\, x\ s(x) 
\ .
\label{F2decompo}
\end{eqnarray} 

We choose the following ansatz for the input valence distribution 
\begin{equation}
 x\, v(x)  =  N_v\, x^{a_v} (1-x)^{b_v} \left[ 1 + C_v \sqrt{x} \right] 
\label{valansatz}
\end{equation}
and analogous forms for the sea and gluon distributions. Note that, 
since we fix the normalization of the hadronic input distributions 
using VMD, we can exploit two constraints, namely the number of 
valence quarks and the momentum sum rule 
\begin{equation}
 \int_0^1 {\d} x\, 2 v(x)  =  2
\quad ; \quad 
 \int_0^1 {\d} x\, \sum_a x\ f_a(x)  =  1
\ .
\label{sumrule}
\end{equation}

\subsection{Direct contribution}

In contrast to the proton structure function $F_2^p$, 
the photon structure function $F_2^\gamma$ receives a direct 
contribution beyond LO accuracy, symbolically 
\begin{equation}
  F_2^{\gamma}(x,Q^2) = \sum_{\q} e_{\q}^2 
  \left[ f_{\q}^\gamma + f_{\qbar}^\gamma 
  \right] \otimes C_{\q} + f_{\g}^\gamma \otimes C_{\g} + C^\gamma
\ .
\label{NLOF2}
\end{equation}
(Note that summation over $\q$ does not automatically include
antiquarks as well; the additional $\qbar$ contribution is reflected 
in a factor 2 in many of the subsequent formulae.) 
In fact, at the relatively large $x$ values currently probed, the
inclusion of the direct term
\begin{equation}
  C^\gamma(x) = 2\ \frac{\alphaem}{2\pi}\, 3\, \sum_{\q}\,  e_{\q}^4\ 
  x\ \left\{ \left[ x^2 + (1-x)^2 \right] \ln \frac{1-x}{x} + 
   8x(1-x) -1 \right\}
\label{naiveCgamma}
\end{equation} 
is much more important than the modification in the $Q^2$ evolution 
when going from LO to NLO. 
Inclusion of $C^\gamma$ defines the 
$\overline{\mathrm{MS}}$ scheme, while $C^\gamma = 0$ defines the DIS 
scheme. At any order in perturbation theory, the theoretical 
prediction for $F_2$ is scheme-dependent. 
This factorization-scheme dependence is reduced when including 
higher-order contributions (i.e.\ when going from LO to NLO to NNLO etc.). 
Many cross sections in $\gamma p$ and $\gamma\gamma$ are still 
calculated in LO accuracy. 
In order to investigate the factorization-scheme dependence associated 
with a LO calculation we shall fit PDFs in LO for both schemes 
\begin{eqnarray}
  \mathrm{DIS:} & & F_2^\gamma(x,Q^2) = 2 \sum_{\q} 
     e_{\q}^2 f_{\q}^\gamma(x,Q^2) 
\nonumber\\
  \overline{\mathrm{MS}}: & & 
   F_2^\gamma(x,Q^2) = 2 \sum_{\q} e_{\q}^2 f_{\q}^\gamma(x,Q^2) + 
   C^\gamma(x)
\ ,
\label{DISvsMS}
\end{eqnarray}
although formally $C^\gamma$ is of NLO. However, we only take that part
of $C^\gamma$ that is universal \cite{AFG}, namely
\begin{equation}
  C^\gamma(x) = 2\ \frac{\alphaem}{2\pi}\, 3\, \sum_{\q}\,  e_{\q}^4\ x\ 
  \left\{ \left[ x^2 + (1-x)^2 \right] \ln \frac{1}{x} + 
   6x(1-x) -1 \right\} ~.
\label{trueCgamma}
\end{equation} 
This modified expression follows from a careful analysis of the 
ladder diagrams,
but can also be obtained by simply imposing a $t$ (and $u$) cut $t_0$ 
on the box diagram 
to separate the perturbative (point-like) from the non-perturbative 
(hadronic) parts resulting in 
\begin{eqnarray}
\lefteqn{  F_2^{\gamma,\mathrm{box}}(x,Q^2) = }
\nonumber \\ & & 
  2\ \frac{\alphaem}{2\pi}\, 3\, \sum_{\q}\,  e_{\q}^4\ x\ 
  \left\{ \left[ x^2 + (1-x)^2 \right] 
  \ln \frac{t_{\mrm{max}}-t_0}{t_0} 
   + \left[ 6x(1-x) -1 \right] 
  \left(1 - \frac{2t_0}{t_{\mrm{max}}} \right)
   \right\}
\label{boxCgamma}
\end{eqnarray} 
with $t_{\mrm{max}}=Q^2/x$. Upon dropping the higher-twist terms 
($\propto 1/Q^2$), eq.~(\ref{boxCgamma}) yields eq.~(\ref{trueCgamma}). 
There is no point 
in keeping the higher-twist contribution in $C^\gamma$ when neglecting 
the further (unknown) higher-twist contributions to $F_2^\gamma$.  
Since charm (and bottom) are included by the Bethe-Heitler formulae,
they are not affected by the considerations in this section. Hence the
quark sum here only runs over $\u$, $\d$ and $\s$ quarks.

\subsection{Low-$Q_0$ fit}

The $F_2^{\gamma}$ data are not yet precise enough for 
the QCD scale parameter $\Lambda$ to be fitted. 
Therefore we fix (one-loop) $\Lambda(n_f=3) = 230\,$MeV 
corresponding to $\Lambda(n_f=4) = 200\,$MeV. 
Moreover, for $K=1$, $Q_0 = 0.6\,$GeV is fixed from our $\gamma\p$ 
analysis. In the DIS scheme, the simplest choice, a valence-like 
input ($N_s = 0$ and 
$g(x) \propto v(x)$ with $v(x=0)=0$), results in a reasonable $\chi^2$ 
only for a considerably smaller $Q_0 \sim 0.4\,$GeV. Therefore 
we allow for sea-like input distributions. As the data do not 
require non-zero values for the parameters $a_s$ and $C_i$, 
cf. eq.~(\ref{valansatz}), we simply put these parameters equal to zero.
The $b_g$ and $\b_s$ are poorly constrained, and have
therefore been assumed to be given by $b_g = b_v +1$,
$b_s = b_v + 3$. Finally, $a_g = a_v/2$ is not inconsistent
with the data and has been fixed.
Then we find in the DIS scheme
\begin{eqnarray}
  \mrm{set} \; \mrm{SaS}\ \mrm{1D} & &  \mrm{input:} \nonumber \\
  \mathrm{DIS:} & & \quad Q_0 = 0.6\,\mathrm{GeV} \quad , \quad 
           \Lambda = 0.2\,\mathrm{GeV}
\nonumber\\
  x\ v(x) & = & 1.294 x^{0.80} (1-x)^{0.76} 
\nonumber\\
  x\ s(x) & = & 0.100 (1-x)^{3.76}
\nonumber\\
 x\ g(x) & = & 1.273 x^{0.40} (1-x)^{1.76}
\label{fitresult}
\end{eqnarray} 
at a $\chi^2$ of $141$ for $71$ $F_2^\gamma$ data points. 
Note that we have 
included also the low-$Q^2$ points ($Q^2 = 0.71\,$GeV$^2$) in the fit.  
At the input scale 
$Q=Q_0$, the photon momentum is split according to about $5:1:2$ 
between the (two) valence quarks, the (six) sea quarks, and the gluon. 

In the $\overline{\mathrm{MS}}$ scheme the fit favours a small sea and 
we find
\begin{eqnarray}
  \mrm{set} \; \mrm{SaS}\ \mrm{1M} & &  \mrm{input:} \nonumber \\
  \overline{\mathrm{MS}}: & & \quad Q_0 = 0.6\,\mathrm{GeV} \quad , \quad 
           \Lambda = 0.2\,\mathrm{GeV}
\nonumber\\
  x\ v(x) & = & 0.8477 x^{0.51} (1-x)^{1.37} 
\nonumber\\
  x\ s(x) & = & 0
\nonumber\\
 x\ g(x) & = & 3.42 x^{0.255} (1-x)^{2.37}
\label{fit2result}
\end{eqnarray} 
at a $\chi^2$ of $136$,  
where the photon momentum is now split according to about $7:0:13$ 
at the input scale $Q=Q_0$. 
Both $Q^2$-evolved PDFs are parametrized as 
\begin{equation}
x f_{a}^{\gamma,\mrm{VMD}}(x,s) = c_1^a \, 
x^{\displaystyle c_2^a} \, 
(1-x)^{\displaystyle c_3^a} 
\left[- \ln x\right]^{\displaystyle c_4^a} 
+  c_5^a \, x^{\displaystyle c_6^a} \, 
(1-x)^{\displaystyle c_7^a} 
\quad a = \mrm{q,val},~\mrm{q,sea},~\g
\ , 
\label{hadoneparam}
\end{equation} 
and the coefficients $c_i$ are given in tables~\ref{hadonetable} 
and~\ref{hadtwotable}, respectively. The coefficients $c_5$--$c_7$
are only listed in those cases where the second term was included.

\begin{table}
\begin{center}
\begin{tabular}{|c|r|r|r|r|r|r|r|r|}
\cline{2-9} 
\multicolumn{1}{c|}{}& $a$ & $b$ & $c$ & $d$ & $e$ & $f$ & $g$ & $\kappa$
\\ \hline
$c_1^{\mathrm{val}}$ & $1.294$ & -- & -- & -- & $0.252$ & $3.079$ & -- & -- 
\\ \hline
$c_2^{\mathrm{val}}$ & $0.80$ & $- 0.13$ & -- & -- & -- & -- & --  & -- 
\\ \hline
$c_3^{\mathrm{val}}$ & $0.76$ & $0.667$ & -- & -- & -- & -- & --  & -- 
\\ \hline
$c_4^{\mathrm{val}}$ & -- & $2.0$ & -- & -- & -- & -- & --  & -- 
\\ \hline
$c_1^{\mathrm{sea}}$ & $0.1$ & -- & $- 0.397$ & $1.121$ & -- & $5.61$
      & $5.26$  & -- 
\\ \hline
$c_2^{\mathrm{sea}}$ & -- & -- & $-7.32$ & -- & -- & $10.3$ & --  & --
\\ \hline
$c_3^{\mathrm{sea}}$ & $3.76$ & $15.0$ & $12.0$ & -- & $4.0$ & -- & --  
   & --
\\ \hline
$c_4^{\mathrm{sea}}$ & -- & -- & -- & -- & -- & -- & --  & --
\\ \hline
$c_1^{\mathrm{g}}$ & -- & $7.90$ & -- & -- & $5.50$ & -- & --  & $5.16$
\\ \hline
$c_2^{\mathrm{g}}$ & -- & $-1.9$ & -- & -- & $3.6$ & -- & --  & --
\\ \hline
$c_3^{\mathrm{g}}$ & $1.3$ & -- & -- & -- & -- & -- & --  & --
\\ \hline
$c_4^{\mathrm{g}}$ & $0.5$ & $3.0$ & -- & -- & -- & -- & --  & --
\\ \hline
$c_5^{\mathrm{g}}$ & $1.273$ & -- & -- & -- & -- & -- & --  & $10.0$
\\ \hline
$c_6^{\mathrm{g}}$ & $0.4$ & -- & -- & -- & -- & -- & --  & --
\\ \hline
$c_7^{\mathrm{g}}$ & $1.76$ & $3.0$ & -- & -- & -- & -- & --  & --
\\ \hline
\end{tabular}
\end{center}
\caption{Coefficients of the hadronic distributions 
evolved from (\protect\ref{fitresult}), 
parametrized in the form (\protect\ref{hadoneparam}), where 
$c_i = \exp(-\kappa s) (a+bs + cs^2 + d s^3)/(1+es+fs^2+gs^3)$.}
\label{hadonetable}
\end{table}

\begin{table}
\begin{center}
\begin{tabular}{|c|r|r|r|r|r|r|r|r|}
\cline{2-9} 
\multicolumn{1}{c|}{}& $a$ & $b$ & $c$ & $d$ & $e$ & $f$ & $g$ & $\kappa$
\\ \hline
$c_1^{\mathrm{val}}$ & $0.8477$ & -- & -- & -- & $1.37$ & $2.18$ 
   & $3.73$ & -- 
\\ \hline
$c_2^{\mathrm{val}}$ & $0.51$ & $0.21$ & -- & -- & -- & -- & --  & -- 
\\ \hline
$c_3^{\mathrm{val}}$ & $1.37$ & -- & -- & -- & -- & -- & --  & -- 
\\ \hline
$c_4^{\mathrm{val}}$ & -- & $2.667$ & -- & -- & -- & -- & --  & -- 
\\ \hline
$c_1^{\mathrm{sea}}$ & -- & $0.842$ & -- & -- & $21.3$ & $-33.2$
      & $229.$  & -- 
\\ \hline
$c_2^{\mathrm{sea}}$ & $0.13$ & $-2.90$ & -- & -- & $5.44$ & -- & --  & --
\\ \hline
$c_3^{\mathrm{sea}}$ & $3.45$ & $0.5$ & -- & -- & -- & -- & --  & --
\\ \hline
$c_4^{\mathrm{sea}}$ & $2.8$ & -- & -- & -- & -- & -- & --  & --
\\ \hline
$c_1^{\mathrm{g}}$ & -- & $24.0$ & -- & -- & $9.6$ & $0.92$ & $14.34$ &
   $5.94$
\\ \hline
$c_2^{\mathrm{g}}$ & $-0.013$ & $-1.8$ & -- & -- & $3.14$ & -- & --  & --
\\ \hline
$c_3^{\mathrm{g}}$ & $2.37$ & $0.4$ & -- & -- & -- & -- & --  & --
\\ \hline
$c_4^{\mathrm{g}}$ & $0.32$ & $3.6$ & -- & -- & -- & -- & --  & --
\\ \hline
$c_5^{\mathrm{g}}$ & $3.42$ & -- & -- & -- & -- & -- & --  & $12.0$
\\ \hline
$c_6^{\mathrm{g}}$ & $0.255$ & -- & -- & -- & -- & -- & --  & --
\\ \hline
$c_7^{\mathrm{g}}$ & $2.37$ & $3.0$ & -- & -- & -- & -- & --  & --
\\ \hline
\end{tabular}
\end{center}
\caption{Coefficients of the hadronic distributions 
evolved from (\protect\ref{fit2result}), 
parametrized in the form (\protect\ref{hadoneparam}), where 
$c_i = \exp(-\kappa s) (a+bs + cs^2 + d s^3)/(1+es+fs^2+gs^3)$.}
\label{hadtwotable}
\end{table}

\begin{figure}
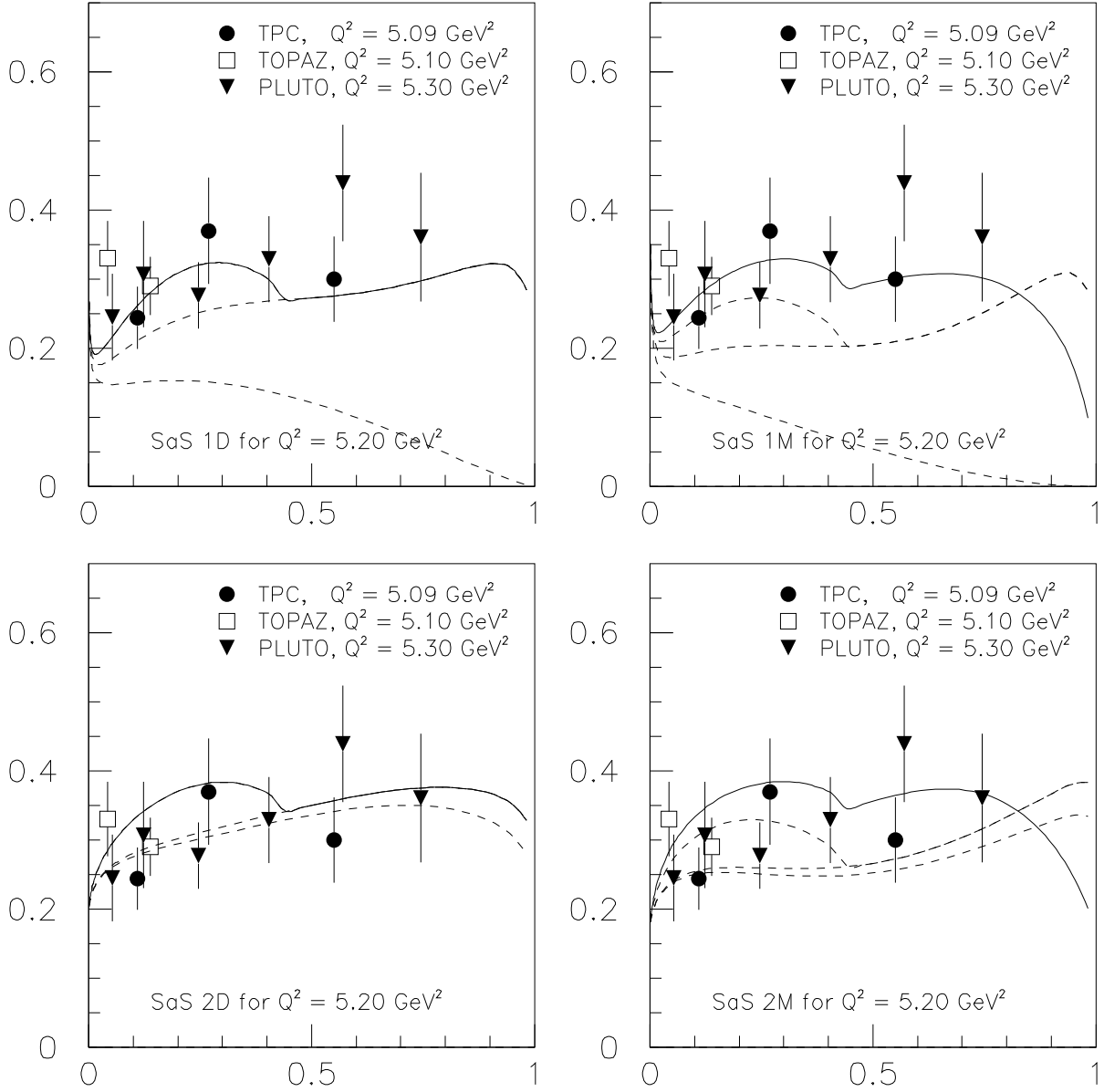

\caption{Subdivision of the full $F_2^{\gamma}$ parametrization by
component, compared with data at $Q^2 \approx 5.2$~GeV$^2$.
The total $F_2^{\gamma}$ is shown by the full curve. 
The lowest dashed curve gives the VMD contribution, and the next 
lowest the sum of VMD and anomalous ones. The third dashed curve,
which coincides with the full curve for the DIS fits, gives
the sum of VMD, anomalous and Beithe--Heitler terms. For the 
$\overline{\mrm{MS}}$ fits, the full curve additionally contains the
contribution of the $C^{\gamma}$ term. Note that this last term is 
negative at large $x$.}
\label{figtwo}
\end{figure}

Figure~\ref{figone} shows $F_2^{\gamma}(x,Q^2)$ data compared with the 
predictions of our sets. As explained above, these consist of
a VMD part (different for the two sets), an anomalous part
for $\u$, $\d$ and $\s$ quarks (in common), Bethe--Heitler terms
for $\c$ and $\b$ production (also in common), and, for set 1M only,
the $C^{\gamma}$ term for $\u$, $\d$ and $\s$ quarks. 
This subdivision is illustrated in Fig.~\ref{figtwo}, for one specific
$Q^2$ scale. With the rather low value of $Q_0$, the anomalous 
contribution quickly becomes the dominant one. At higher $Q^2$ and 
smaller $x$, also the charm contribution
is important. In the $\overline{\mathrm{MS}}$ scheme the hadronic 
distributions vanish faster with $x$ than in the DIS scheme, both
since the $C^\gamma$ term is negative at large $x$ and since the 
$(1-x)$-power is larger.

A reasonable description is obtained for all $Q^2$,  
although there might be a slight incompatibility between the 
low- and high-$Q^2$ data, which is also reflected in the 
not quite optimal $\chi^2$. This could indicate that  
higher-twist contributions are not negligible at low $Q^2$. 
One can imagine a situation in which 
higher-twist contributions significantly affect $F_2$, but 
where the leading-twist evolution of PDFs is still valid 
down to $Q_0 = 0.6\,$GeV. In fact, including only the 
$F_2^\gamma(x,Q^2)$ data above $Q^2 = 4\,$GeV$^2$ improves the fit 
to a $\chi^2$ of $60$ ($78$) in the $\overline{\mathrm{MS}}$ 
(DIS) scheme at now $55$ data points. However, in both cases a
very large sea component is needed. Therefore we do not provide 
such a parametrization but rather one where we also take 
a large $Q_0=2\,$GeV. 

The distributions (\ref{fitresult}) vanish as $x \rightarrow 1$, 
as do truly hadronic distributions. In order to check whether 
the data prefer a harder (point-like) component we added a term 
$N_v d_v x$ to eq.~(\ref{valansatz}) but found no improvement in $\chi^2$. 
We checked also the normalization of the VMD couplings by allowing 
the values $g_i^2$ in table~\ref{coeff} to be multiplied by an overall 
constant $K$. Interestingly enough, 
the data do not really require values for 
the VMD couplings that deviate notably from their values used 
in the description of $\gamma\p$ cross sections. The fit gives 
$K=1.17$ with only a slightly better $\chi^2$ ($136$ compared to $141$). 
This $17\%$ deviation of $K$ from unity 
is within the range of uncertainty of 
the $g_i^2$. For example, using $f_{\rho}^2/4\pi=2.02$, as extracted from 
the leptonic width alone, gives a $9\%$ increase compared to the value 
we use, namely $2.20$, which is the geometrical mean between
this ``leptonic" value and the value extracted from $\gamma\p 
\rightarrow \rho^0 \p$. 

\subsection{High-$Q_0$ fit} 

\begin{figure}
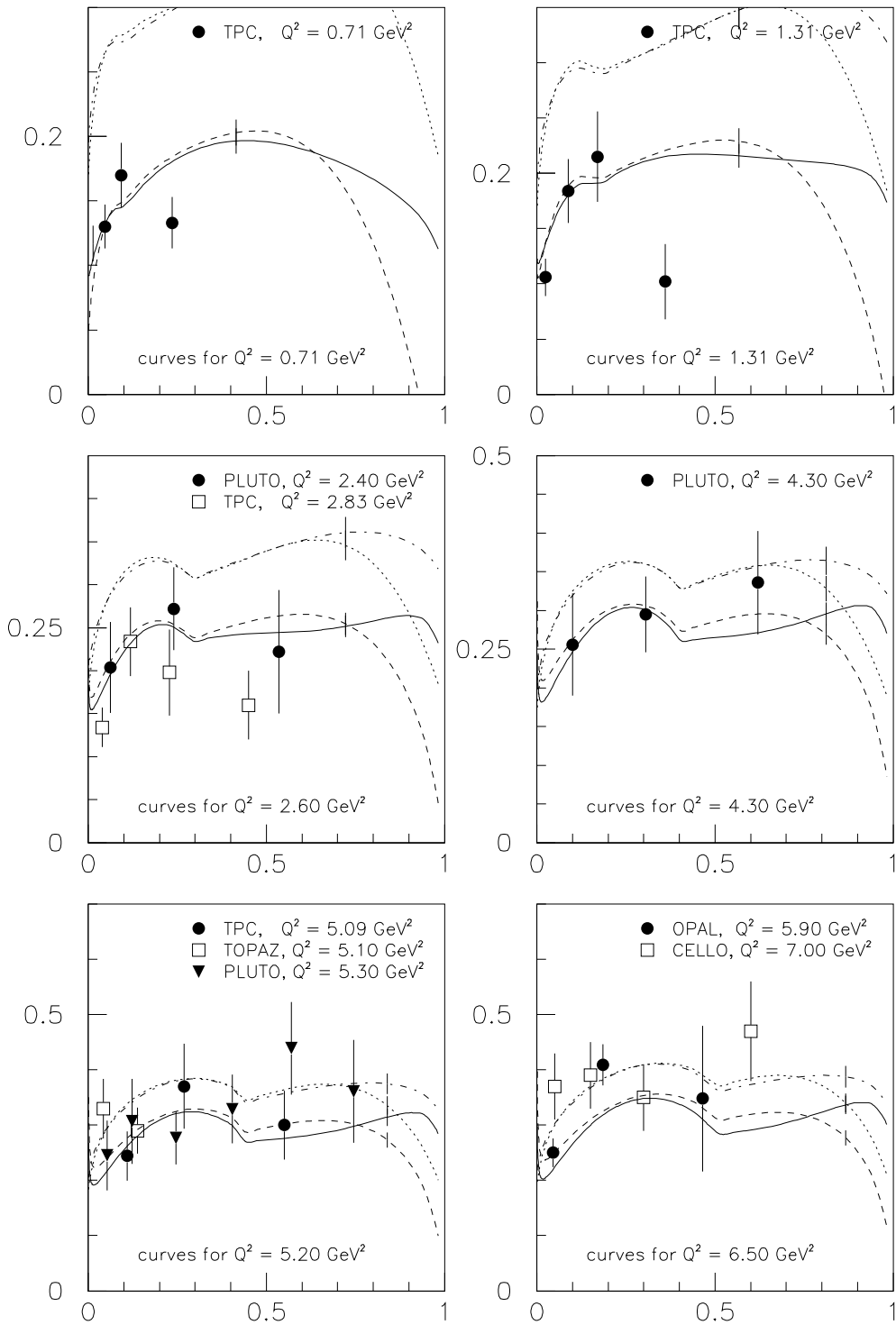

\caption{Comparison of the fits with the available 
 \protect\cite{JADE,PLUTO,TASSO,TPC,AMY,CELLO,TOPAZ,OPAL} 
 $F_2^\gamma$ data. The full curve is for set SaS 1D, dashed for
 SaS 1M, dash-dotted SaS 2D and dotted SaS 2M. Data at nearby
$Q^2$ values are shown in the same frame, and the theoretical
curves are evaluated at some average $Q^2$ value of that range.
The small vertical bars through the curves indicate the $x$ value
where $W=1$~GeV. The behaviour to the right of this point should
be taken as indicative only, with the possibility of large
higher-twist contributions.}
\label{figone}
\end{figure}

Alternatively we fit input distributions starting at 
a ``typical" deep-inelastic $Q_0$ value, namely $Q_0=2\,$GeV. 
Fixing again $\Lambda_4$ at $200\,$MeV, as well as 
$a_s = a_g = C_i = 0$, $b_s = 4$, $b_g = 2$, 
we find in the DIS scheme 
$K=2.422$ 
(corresponding to additionally allowed vector mesons between $0.6\,$GeV 
and $2\,$GeV) and 
\begin{eqnarray}
  \mrm{set} \; \mrm{SaS}\ \mrm{2D} & &  \mrm{input:} \nonumber \\
  \mathrm{DIS:} & & \quad Q_0 = 2\,\mathrm{GeV} \quad , \quad 
           \Lambda = 0.2\,\mathrm{GeV} \quad , \quad  K=2.422
\nonumber\\
  K x\ v(x) & = & 1.00 \left[ x^{0.46} (1-x)^{0.64} + 0.76 x \right]
\nonumber\\
  K x\ s(x) & = & 0.242 (1-x)^{4}
\nonumber\\
 K x\ g(x) & = & 1.925 (1-x)^{2}
\label{highq2fit}
\end{eqnarray} 
with $\chi^2 = 59$ for $55$ $F_2^\gamma$ data points 
above $Q^2 = 4\,$GeV$^2$. In this case the 
photon momentum is split as about $5:1:2$ between valence quarks, 
sea quarks, 
and gluons, where the ``hadronic" term and the ``hard" term in 
eq.~(\ref{highq2fit}) contribute each about $50\%$ to the valence 
momentum. As can be seen from Fig.~\ref{figone}, this fit describes 
the high-$Q^2$ 
data better than the fit starting at $Q_0=0.6\,$GeV, however, at the 
expense of not accommodating the low-$Q^2$ data. 

As in the case of the low-$Q^2$ fits, 
the fit in the $\overline{\mathrm{MS}}$ scheme again favours softer 
distribution functions and we find
\begin{eqnarray}
  \mrm{set} \; \mrm{SaS}\ \mrm{2M} & &  \mrm{input:} \nonumber \\
  \overline{\mathrm{MS}}: & & \quad Q_0 = 2\,\mathrm{GeV} \quad , \quad 
           \Lambda = 0.2\,\mathrm{GeV} \quad , \quad K=2.094
\nonumber\\
  K x\ v(x) & = & 1.168 \left[ x^{0.50} (1-x)^{2.60} + 0.826 x \right]
\nonumber\\
  K x\ s(x) & = & 0.209 (1-x)^{4}
\nonumber\\
 K x\ g(x) & = & 1.808 (1-x)^{2}
\label{highq2MS}
\end{eqnarray} 
with the same $\chi^2$ and a slightly larger gluon-momentum fraction 
($29\%$ compared to $27\%$). The comparison with data and the low-$Q^2$ fit 
is given in Fig.~\ref{figone}. 

The breakdown of the two sets component by component is again shown in
Fig.~\ref{figtwo}. Owing to the larger $Q_0$ scale, the anomalous
component is small at the lower $Q^2$ range. Note that the 
Bethe--Heitler contribution to $\c\cbar$ production remains unchanged, 
since we do not include an increased amount (or, indeed, any amount at 
all) of $\J/\psi$ production in the VMD component and therefore are
allowed to maintain the same $m_{\c} = 1.3$~GeV as was used for the
low-$Q_0$ fits.  

The distributions have again been parametrized according to 
eq.~(\ref{hadoneparam}), with coefficients given in tables
\ref{hadthreetable} and
\ref{hadfourtable}, respectively.

\begin{table}
\begin{center}
\begin{tabular}{|c|r|r|r|r|r|r|r|r|}
\cline{2-9} 
\multicolumn{1}{c|}{}& $a$ & $b$ & $c$ & $d$ & $e$ & $f$ & $g$ & $\kappa$
\\ \hline
$c_1^{\mathrm{val}}$ & $1.0$ & $0.186$ & -- & -- & $-0.209$ & 
$1.495$ & -- & -- 
\\ \hline
$c_2^{\mathrm{val}}$ & $0.46$ & $0.25$ & -- & -- & -- & -- & --  & -- 
\\ \hline
$c_3^{\mathrm{val}}$ & $0.64$ & $0.14$ & $5.0$ & -- & $1.0$ & -- & --  & -- 
\\ \hline
$c_4^{\mathrm{val}}$ & -- & $1.9$ & -- & -- & -- & -- & --  & -- 
\\ \hline
$c_5^{\mathrm{val}}$ & $0.76$ & $0.4$ & -- & -- & -- & -- & --  & --
\\ \hline
$c_6^{\mathrm{val}}$ & $1.0$ & -- & -- & -- & -- & -- & --  & --
\\ \hline
$c_7^{\mathrm{val}}$ & -- & $2.667$ & -- & -- & -- & -- & --  & --
\\ \hline
$c_1^{\mathrm{sea}}$ & $0.242$ & $-0.252$ & $1.19$ & -- & $-0.607$ & 
      $21.95$ & -- & -- 
\\ \hline
$c_2^{\mathrm{sea}}$ & -- & -- & $-12.1$ & -- & $2.62$ & $16.7$ & --  & --
\\ \hline
$c_3^{\mathrm{sea}}$ & $4.0$ & -- & -- & -- & -- & -- & --  & --
\\ \hline
$c_4^{\mathrm{sea}}$ & -- & $1.0$ & -- & -- & -- & -- & --  & --
\\ \hline
$c_1^{\mathrm{g}}$ & $1.925$ & $5.55$ & $147.$ & -- & $-3.59$ & $3.32$ 
    & --  & $18.67$
\\ \hline
$c_2^{\mathrm{g}}$ & -- & $-5.81$ & $-5.34$ & -- & $29.0$ & $-4.26$ 
   & --  & --
\\ \hline
$c_3^{\mathrm{g}}$ & $2.0$ & $-5.9$ & -- & -- & $1.7$ & -- & --  & --
\\ \hline
$c_4^{\mathrm{g}}$ & -- & $9.3$ & -- & -- & $1.7$ & -- & --  & --
\\ \hline 
\end{tabular}
\end{center}
\caption{Coefficients of the hadronic distributions 
evolved from (\protect\ref{highq2fit}), 
parametrized in the form (\protect\ref{hadoneparam}), where 
$c_i = \exp(-\kappa s) (a+bs + cs^2 + d s^3)/(1+es+fs^2+gs^3)$.}
\label{hadthreetable}
\end{table}

\begin{table}
\begin{center}
\begin{tabular}{|c|r|r|r|r|r|r|r|r|}
\cline{2-9} 
\multicolumn{1}{c|}{}& $a$ & $b$ & $c$ & $d$ & $e$ & $f$ & $g$ & $\kappa$
\\ \hline
$c_1^{\mathrm{val}}$ & $1.168$ & $1.771$ & $29.35$ & -- & -- & -- & -- 
    & $5.776$ 
\\ \hline
$c_2^{\mathrm{val}}$ & $0.5$ & $0.208$ & -- & -- & $-0.794$ & $1.516$ 
   & --  & -- 
\\ \hline
$c_3^{\mathrm{val}}$ & $2.6$ & $7.6$ & -- & -- & $5.0$ & -- & --  & -- 
\\ \hline
$c_4^{\mathrm{val}}$ & -- & $5.15$ & -- & -- & $2.0$ & -- & --  & -- 
\\ \hline
$c_5^{\mathrm{val}}$ & $0.965$ & $22.35$ & -- & -- & $18.4$ & -- & --  & --
\\ \hline
$c_6^{\mathrm{val}}$ & $1.0$ & -- & -- & -- & -- & -- & --  & --
\\ \hline
$c_7^{\mathrm{val}}$ & -- & $2.667$ & -- & -- & -- & -- & --  & --
\\ \hline
$c_1^{\mathrm{sea}}$ & $0.209$ & -- & $0.644$ & -- & $0.319$ & $17.6$
      & -- & -- 
\\ \hline
$c_2^{\mathrm{sea}}$ & $-0.373$ & $-7.71$ & -- & -- & $0.815$ & $11.0$ &
     --  & --
\\ \hline
$c_3^{\mathrm{sea}}$ & $4.0$ & $1.0$ & -- & -- & -- & -- & --  & --
\\ \hline
$c_4^{\mathrm{sea}}$ & -- & $0.45$ & -- & -- & -- & -- & --  & --
\\ \hline
$c_1^{\mathrm{g}}$ & $1.808$ & $29.9$ & -- & -- & $26.4$ & -- & -- &
   $5.28$
\\ \hline
$c_2^{\mathrm{g}}$ & -- & $-5.35$ & $-10.11$ & -- & $31.71$ & -- & --  & --
\\ \hline
$c_3^{\mathrm{g}}$ & $2.0$ & $-7.3$ & $4.0$ & -- & $2.5$ & -- & --  & --
\\ \hline
$c_4^{\mathrm{g}}$ & -- & $10.9$ & -- & -- & $2.5$ & -- & --  & --
\\ \hline
\end{tabular}
\end{center}
\caption{Coefficients of the hadronic distributions 
evolved from (\protect\ref{highq2MS}), 
parametrized in the form (\protect\ref{hadoneparam}), where 
$c_i = \exp(-\kappa s) (a+bs + cs^2 + d s^3)/(1+es+fs^2+gs^3)$.}
\label{hadfourtable}
\end{table}

\begin{figure}
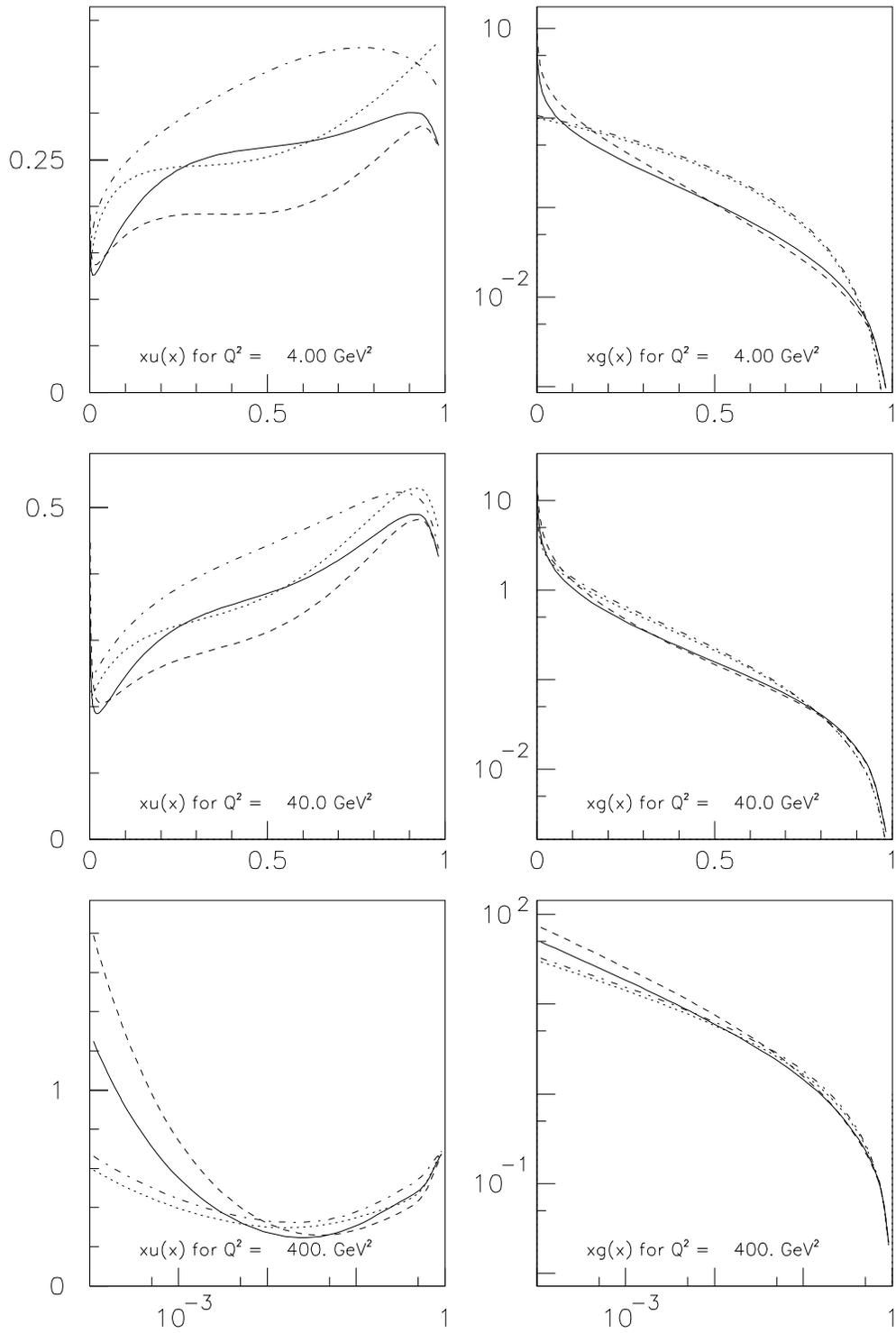

\caption{Comparison of parametrizations of the $\u$-quark and gluon PDFs 
at $Q^2 = 4, 40$ and $400$~GeV$^2$. Full curve is for set SaS 1D, 
dashed for SaS 1M, dash-dotted SaS 2D and dotted SaS 2M, as in 
Fig.~\protect\ref{figone}. Note that a logarithmic $y$ scale is
used for the gluon distribution. For $Q^2 = 400$~GeV$^2$ also the
$x$ axis is logarithmic, so as better to show differences in the
small-$x$ behaviour.}
\label{figthree}
\end{figure}

The current uncertainties in the PDFs of the photon are displayed 
in Fig.~\ref{figthree}, which shows the $\u$-quark and gluon PDFs. 
For the quark distribution, the high-$Q^2$ distributions SaS 2D and 2M 
are larger than the low-$Q^2$ ones 1D and 1M, simply reflecting 
the corresponding hierarchy in $F_2^\gamma$. At small $x$ the opposite 
behaviour holds true since the small-$x$ rise increases with
the length of the $Q^2$ evolution. Hence the low-$Q^2$ parametrizations 
also yield larger gluon distributions at small $x$.  

\begin{figure}
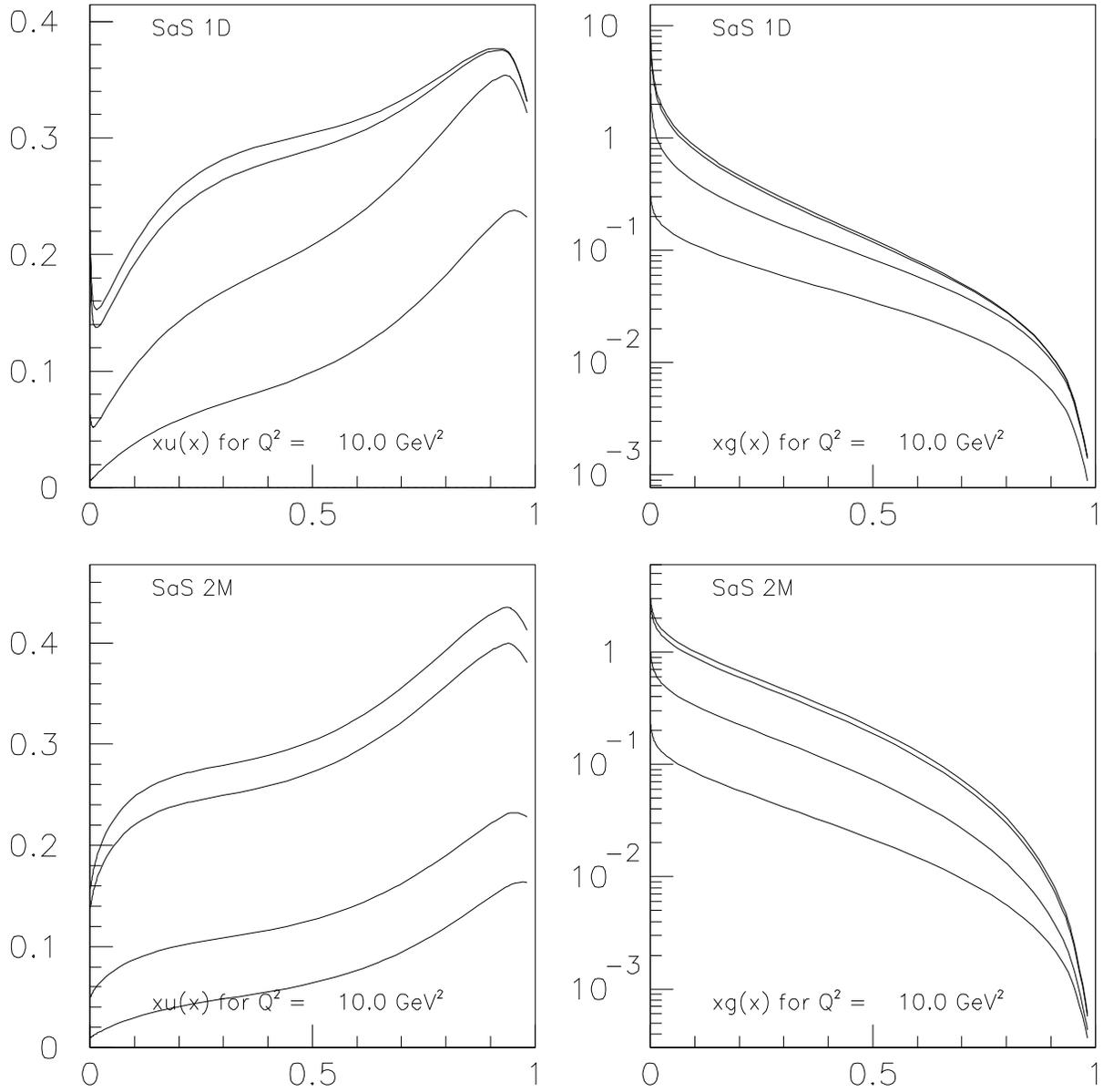

\caption{Target-mass ($P^2$) dependence of the $\u$-quark and gluon PDFs,
for $Q^2 = 10$~GeV$^2$. From top to bottom: $\protect\sqrt{P^2}= 0$, 
$0.2\protect\,$GeV, $0.7\protect\,$GeV, and $1.4\protect\,$GeV.
The top two frames are for set SaS 1D, and the bottom two 
for set SaS 2M. The remaining two look similar qualitatively.}
\label{figfour}
\end{figure}

The effect of a non-zero target virtuality is displayed in 
Fig.~\ref{figfour}. Increasing $P^2$ not only lowers the normalization 
but also changes the shape of the distribution. As expected, the 
gluon distribution vanishes faster than the quark distribution 
as $P^2 \rightarrow Q^2$. 

\subsection{The TPC/$2\gamma$ parametrization and comparison with FKP} 

\label{FKPsec}

Experiments often use the TPC/$2\gamma$ parametrization \cite{TPC} of 
$F_2^\gamma(x,Q_0^2)$  
determined at $Q_0^2 =0.71\,$GeV$^2$  
to estimate the hadronic component of $F_2^\gamma(x,Q^2)$ 
\begin{eqnarray}
  \frac{1}{\alphaem}\, F_2^{\gamma,\mathrm{VMD}}(x,Q^2) & \approx &  
  \frac{1}{\alphaem}\, F_2^{\gamma}(x,Q_0^2)|_{\mathrm{TPC}/2\gamma} 
\label{TPCF2}
\\ \nonumber  
 & \equiv & (0.22\pm 0.01)\, x^{0.31\pm 0.02}\, (1-x)^{0.95}
 + (0.06\pm 0.01)\, (1-x)^{2.5\pm 1.1}
\ .
\end{eqnarray} 
The full $F_2^\gamma(x,Q^2)$ distribution is then obtained by adding 
the FKP parametrization \cite{FKP} of the anomalous part of 
$F_2^\gamma$ to eq.~(\ref{TPCF2}). The 
only free parameter\footnote{The values of the light-quark masses are 
needed as well, but these are supposed to be known.}
in this aproach is the cutoff scale $p_{T0}$ 
separating the perturbative part of $F_2^\gamma$ from the 
non-perturbative one. Moreover, $p_{T0}$ is often varied 
in the figures of $F_2^\gamma(x,Q^2)$ or 
$\int_{0.3}^{0.8} dx F_2^\gamma(x,Q^2)$ to illustrate the strong 
dependence on this arbitrary scale, implying little or 
no sensitivity of the data to the QCD scale $\Lambda$. 
This approach to compose $F_2^\gamma$ from a hadronic and a 
perturbative part can be criticized on three counts: 
\begin{Enumerate}
\item The hadronic part of $F_2^\gamma$, eq.~(\ref{TPCF2}), is 
  not evolved in $Q^2$.
\item The strong correlation between the scale $p_{T0}$ and the 
 size and shape of the hadronic part is neglected.
\item The FKP formula is a poor parametrization of the anomalous part.
\end{Enumerate}
Let us discuss these points in more detail.

The first point is obvious. In order to investigate 
how well the $F_2^\gamma$ data can be described with eq.~(\ref{TPCF2}), 
one has to consider this equation as the non-perturbative input 
to $F_2^\gamma(x,Q^2)$ at the scale $Q_0^2 = 0.71\,$GeV$^2$, 
whereafter it evolves with $Q^2$. Note that, for the $Q^2$ evolution, the 
input distributions have to be known separately for the valence, sea, 
and gluon distributions. It is sensible to assume that  
these hadronic input distributions can be interpreted within 
generalized vector-meson dominance (GVMD), in which case we can exploit 
eq.~(\ref{sumrule}). Associating the first term of eq.~(\ref{TPCF2}) with 
valence quarks, we need a $K$-factor of $K=1.411$
compared to simple VMD, eq.~(\ref{F2decompo}). This represents  
the contribution of additional vector mesons that have to be included 
in eqs.~(\ref{VMDgammaPDF}) and (\ref{F2decompo})  
because of the larger continuum cutoff $Q_0=0.84\,$GeV 
compared to the $\sim 0.6\,$GeV preferred for simple VMD. 

\begin{figure}
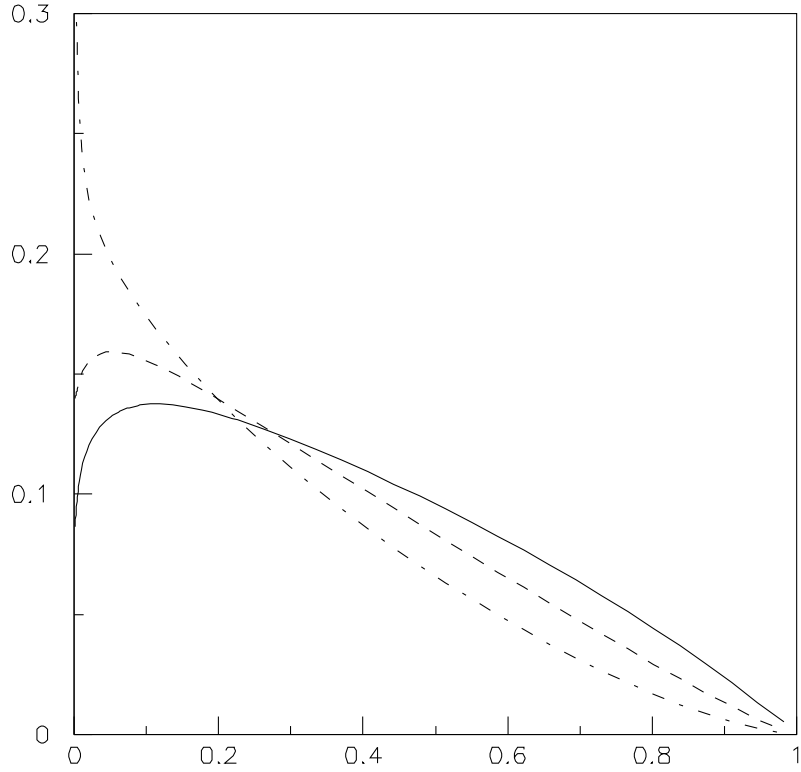

\caption{The hadronic part of $F_2^{\gamma}(x,Q^2)$ 
 as obtained by evolving the TPC/$2\gamma$ parametrization of 
 $F_2^\gamma$. The full curve is the $x$ dependence at the input scale
 $Q^2 = 0.71\,$GeV$^2$, dashed curve at $4\,$GeV$^2$, and 
 dash-dotted curve at $100\,$GeV$^2$.}
\label{TPCevolvefig}
\end{figure}

The only free parameters are then $\Lambda$, $a_g$, and $b_g$. 
The fit to the $F_2^\gamma$ data in the DIS scheme yields 
$\Lambda = 0.116\,$GeV and  
\begin{eqnarray}
  K x\ v(x) & = & 0.57 x^{0.31} (1-x)^{0.95} 
\nonumber\\
  K x\ s(x) & = & 0.082 (1-x)^{2.5}
\nonumber\\
 K x\ g(x) & = & 2.586 x^{0.31} (1-x)^{0.95}
\ .
\label{fitTPC}
\end{eqnarray} 
The photon momentum is approximately 
split as $3:1:6$ between valence quarks, sea quarks, and gluons 
at the input scale.  
In Fig.~\ref{TPCevolvefig}
we show the hadronic part of $F_2^\gamma$ 
at $Q^2=4$ and $100\,$GeV$^2$ as well as at the input scale $0.71\,$GeV$^2$. 
The effect of the $Q^2$ evolution is non-negligible: a clearly visible 
softening at large $x$ and an increase at small $x$ is observed. 
The fit is, however, a rather bad representation of the data, in 
particular at high $Q^2$. This is reflected in a high $\chi^2$ value, 
$\chi^2 = 181$. We therefore do not provide a parametrization for 
the distributions $Q^2$-evolved from eq.~(\ref{TPCF2}). 

Concerning the second point, we recall that, by definition, the 
perturbative part of $F_2^\gamma(x,Q^2)$ vanishes at the 
$Q_0$ (or $p_{T0}$) scale separating it from the non-perturbative 
part of $F_2^\gamma$. Hence, any change in this scale has to be 
accompanied by a corresponding change in
the size and shape of the hadronic input distributions at this scale. 
The correlation is obvious: the larger the $Q_0$, the smaller is the 
range of perturbative evolution, and hence the bigger is the 
non-perturbative input needed. If higher-twist effects can be neglected,
the total photon momentum carried by partons should be independent 
of $Q_0$. Then eqs.~(\ref{Zthreedef}) and (\ref{Zthreevalue}) yield 
the following dependence on $Q_0$ for the 
normalization $K=K(Q_0)$ defined in eq.~(\ref{Kfactor}): 
\begin{eqnarray}
  K(Q'_0) & = & K(Q_0) + \frac{\sum_{\q} e_{\q}^2}
                   {\pi \sum_{V=\rho,\omega,\phi} 4\pi/f_V^2}
  \ln \frac{Q'^2_0}{Q_0^2} 
\nonumber\\
 & \approx & K(Q_0) + 0.770  \ln \frac{Q'_0}{Q_0}
\ .
\label{normrelation}
\end{eqnarray} 

Also a correlation between the separation scale and the $x$-shape  
is to be expected, with harder hadronic input distributions 
required for larger $Q_0$ scales. This could be 
argued on the basis of GVMD, or simply by the fact that parts of 
the (hard) perturbative $\gamma \rightarrow \q\qbar$ splittings 
have to be included in the ``hadronic'' input distributions for 
larger $Q_0$. In any case, 
the separation scale is an arbitrary parameter and hence the result 
for $F_2^\gamma$ {\em must not depend} on its value, provided it is 
changed within a range in which leading-twist perturbative QCD 
describes the physics. As the comparison of our various parametrizations
demonstrates, even the variation of the scale between $0.6\,$GeV and 
$2.0\,$GeV has not a major influence on the final $F_2^\gamma$ result.
(The differences that are there come from them being fits to different
data sets, i.e. data with $Q<2$~GeV included or not. One can do a 
$Q_0=0.6$~GeV fit to $Q>2$~GeV data only, and then obtain close 
agreement with a $Q_0=2$~GeV fit.) 

\begin{figure}
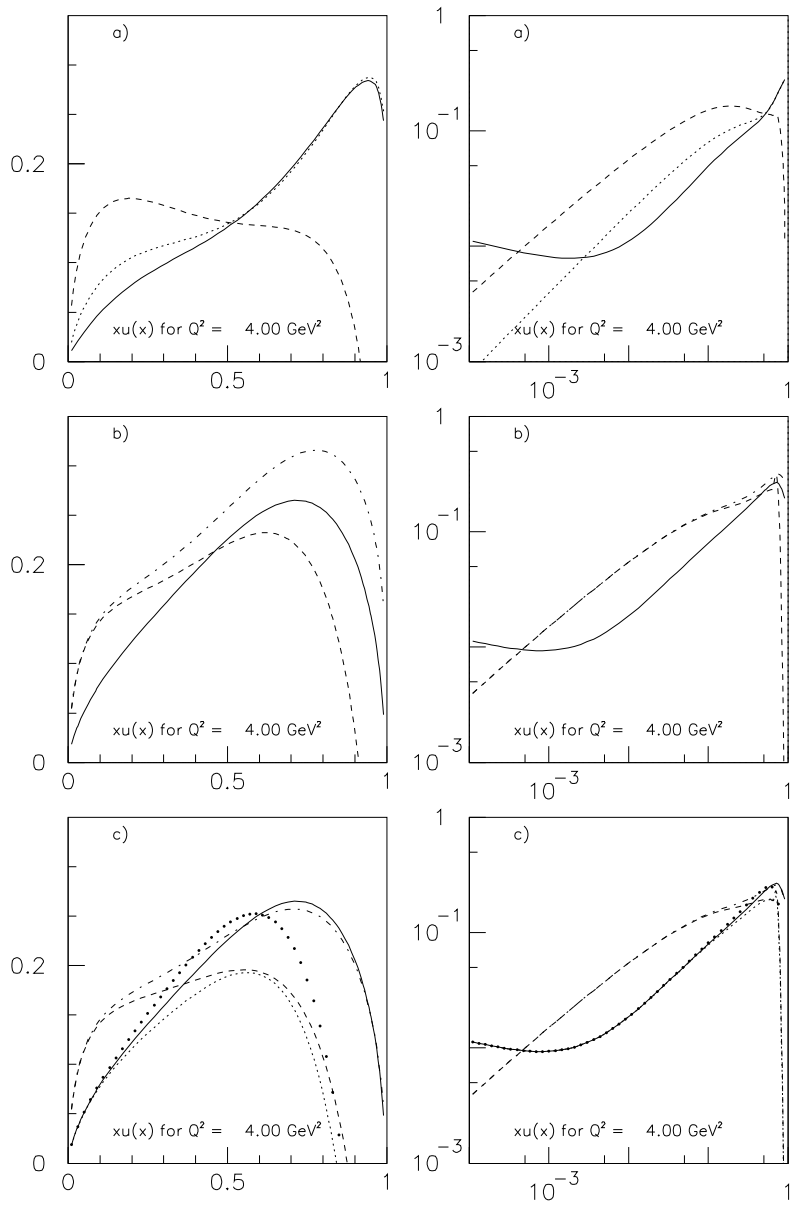

\caption{Comparison between the FKP and our parametrizations of 
the anomalous u-quark distribution as a function of $x$. 
The $x$ and $xu(x)$ values are shown in linear scale to the left 
and in logarithmic scale to the right. Common values are 
$Q^2=4\,$GeV$^2$, $P^2=0$, $n_f=3$, and $\Lambda_3=230\,$MeV; 
additionally $Q_0^2=0.36\,$GeV$^2$ in our parametrization. 
\protect\newline
a) The full curve is our parametrization, 
dashed curve is FKP with $t_{\mrm{max}}=Q^2/x$ and $t_0=Q_0^2/(1-x)$, 
and dotted curve is FKP with $t_{\mrm{max}}=Q^2$ and $t_0=Q_0^2$. 
\protect\newline
b) The full curve is our parametrization plus our direct term 
(\protect\ref{ourdirect}),
dashed curve is FKP with their direct term (\protect\ref{F2FKP}), 
$m_{\q}/p_{T0}=0.3/0.52$, $t_{\mrm{max}}=Q^2/x$ and $t_0=Q_0^2/(1-x)$, 
and dash-dotted curve is FKP as above except that $t_0=Q_0^2$. 
\protect\newline
c) The full curve is our parametrization plus our direct term 
(\protect\ref{ourdirect}), large-dotted curve is our parametrization 
plus the standard direct term (\protect\ref{standarddirect}), dotted 
curve is our parametrization plus the modified direct term 
(\protect\ref{modifieddirect}),
dashed curve is FKP with their direct term (\protect\ref{F2FKP}), 
$m_{\q}/p_{T0}=0./0.6$, $t_{\mrm{max}}=Q^2/x$ and $t_0=Q_0^2/(1-x)$, 
and dash-dotted curve is FKP as above, except that $t_0=Q_0^2$. 
}
\label{figFKP}
\end{figure}

As to point~3 above, a solution of the perturbatively calculable part 
of the quark distribution functions of the photon
has been obtained by FKP \cite{FKP}: 
\begin{equation}
 f_{\q}^{\gamma,\mathrm{anom}}(x,Q^2) =  \frac{\alphaem}{2\pi}\, 
   3\, e_{\q}^2\, \frac{x^2 + (1-x)^2}{x^C + C f(x)}\, 
  \Upsilon \left[ 1 - 
      \left(\frac{\Upsilon_0}{\Upsilon}\right)^{1 + C f(x)} \right]
\ ,
\label{fqFKP}
\end{equation}
where 
\begin{eqnarray}
  C=\frac{8}{33 - 2 N_f} & \qquad &  
       f(x) = 2 \ln\frac{1}{1-x} - x - \frac{1}{2} x^2 
\nonumber\\
  \Upsilon = \ln\frac{t_{\mrm{max}}}{\Lambda^2} & & 
  t_{\mrm{max}} = \frac{Q^2}{x}
\nonumber\\
  \Upsilon_0 = \ln\frac{t_0}{\Lambda^2} & &
   t_0 = \frac{m_{T0}^2}{1-x} 
      \equiv \frac{m_{\q}^2 + p_{T0}^2 + x(1-x) P^2}{1-x} 
\ .
\label{FKPvar}
\end{eqnarray} 
Here $t_0$ is the cutoff in the integration over the virtuality 
of the $t$-channel and $u$-channel quark propagators and   
$P^2$ the target mass (virtuality of the probed photon).  
In Fig.~\ref{figFKP}a 
we compare our anomalous u-quark distribution 
function at $Q^2=4\,$GeV$^2$, $P^2=0$, and $Q_0^2=0.36\,$GeV$^2$
with eq.~(\ref{fqFKP}), where we have taken the FKP value for 
the quark mass ($m_{\q} = 0.3\,$GeV) but adjusted 
$p_{T0}$ such that $m_{T0}^2 = Q_0^2$. 

It might seem surprising that 
the two distributions do not at all agree with each other. 
The difference arises mainly from the inclusion of the $1/x$ and 
$1/(1-x)$ factors in eq.~(\ref{FKPvar}) 
($1/x$ in $t_{\mrm{max}}$ and $\Upsilon$, $1/(1-x)$ in $t_0$ and  
$\Upsilon_0$). 
These terms are of kinematic origin and specific to the 
quark-parton-model result (calculation of the box diagram). 
However, their inclusion in the leading-twist QCD evolution equations 
lacks justification in perturbative QCD:  
the leading-logarithmic approximation sums only the logarithms 
in $Q^2/\Lambda^2$ and makes no statement about the summation of 
$\ln x$ or $\ln(1-x)$ terms 
(beyond, of course, the double-leading-logarithmic approximation). 
In other words: although the factorization scale is arbitrary 
and may well differ from its ``natural" value $Q^2$, it must not 
depend on $x$ in order not to spoil factorization. 

On the other hand, it is satisfying to observe that the FKP 
parametrization agrees almost perfectly with ours at large $x$ 
if we take $t_{\mrm{max}} = Q^2$ and $t_0 = m_{T0}^2$ in 
eq.~(\ref{FKPvar}).  
Differences at small $x$ arise because eq.~(\ref{fqFKP}) 
was derived in the valence approximation, hence $x f_{\q}$ 
given in eq.~(\ref{fqFKP}) vanishes as $x \rightarrow 0$ while 
the full solution gives $x f_{\q}$ rising at small $x$. 

Of course, one could argue that the FKP formula is 
not meant as a parametrization of quark distributions 
but as part of an $F_2^\gamma(x,Q^2)$ parametrization,
in combination with VMD and direct terms.
(After all, it is the FKP $F_2^\gamma$ expression that is being 
used in the experimental analyses.) 
In this context, $\ln x$ and $\ln(1-x)$ 
terms may well be redistributed from the direct term into 
the ``quark distribution" if this provides a convenient approximation 
to the full solution. The FKP prescription for $F_2^\gamma$ is 
\begin{eqnarray}
 F_2^{\gamma,\mathrm{anom}}(x,Q^2) & = & 
  2 \sum_{\q} e_{\q}^2 \, x\ \left\{ \rule[-2ex]{0mm}{5ex}
   f_{\q}^{\gamma,\mathrm{anom}}(x,Q^2) \right.
\nonumber\\ 
   & & \left. +\, \frac{\alphaem}{2\pi} \, 3 \, e_{\q}^2 \left[
    6x(1-x) - 1 + \frac{2x m_{\q}^2 - x(1-x) P^2}{m_{T0}^2} 
    \right] \right\} \ .
\label{F2FKP}
\end{eqnarray} 
Figure~\ref{figFKP}b compares the u-quark component of eq.~(\ref{F2FKP}) 
with our parametrization of the anomalous u-quark distribution 
function, to which we add the u-quark component of the universal 
direct term eq.~(\ref{trueCgamma}), i.e. 
\begin{equation}
  C^\gamma_{\u}(x) = \frac{\alphaem}{2\pi}\, 3\,  
     \left(\frac{2}{3}\right)^2 x\ 
  \left\{ \left[ x^2 + (1-x)^2 \right] \ln \frac{1}{x} + 6x(1-x) -1 \right\}
\ .
\label{ourdirect}
\end{equation} 
The observable discrepancy is, in fact, not unexpected as no $\ln(1-x)$ 
term is included in eq.~(\ref{ourdirect}). However, a discrepancy remains 
even when the $\ln(1-x)$ factor is removed from 
the definition of $\Upsilon_0$ in the FKP formula. 

The reason for this is the extra $m_{\q}^2/m_{T0}^2$ term in 
eq.~(\ref{F2FKP}). Indeed, when we take $m_{\q}=0$ and $t_0=Q_0^2$ 
in eqs.~(\ref{FKPvar}) and (\ref{F2FKP}), we find agreement
at large $x$ between our anomalous-plus-direct u-quark 
distribution function and the FKP one, Fig.~\ref{figFKP}c.
Alternatively we also expect agreement for the comparison of 
the standard FKP result, i.e.\ where $t_0=Q_0^2/(1-x)$, if  
we change from the expression (\ref{ourdirect}) 
for the direct term to the standard direct term, eq.~(\ref{naiveCgamma}): 
\begin{equation}
  C^\gamma_{\u}(x) = \frac{\alphaem}{2\pi}\, 3\,  
     \left(\frac{2}{3}\right)^2 x\ 
  \left\{ \left[ x^2 + (1-x)^2 \right] \ln \frac{1-x}{x} + 
  8x(1-x) -1 \right\}
\ ,
\label{standarddirect}
\end{equation} 
which includes the $\ln(1-x)$ term. This is, however, not the case. 
The difference can be traced back to the difference in the coefficient 
of the $x(1-x)$ term in eq.~(\ref{F2FKP}) and in  
eq.~(\ref{standarddirect}). Indeed, we obtain agreement with the 
standard FKP result at large $x$ if we add to our parametrization of
the $\u$-quark distribution function the following direct term: 
\begin{equation}
  C^\gamma_{\u}(x) = \frac{\alphaem}{2\pi}\, 3\,  
     \left(\frac{2}{3}\right)^2 x\ 
  \left\{ \left[ x^2 + (1-x)^2 \right] \ln \frac{1-x}{x} + 
  6x(1-x) -1 \right\} \ .
\label{modifieddirect}
\end{equation} 

In conclusion, putting $m_{\q}$ equal to zero and changing the 
$6x(1-x)$ term into $8x(1-x)$ in eq.~(\ref{F2FKP}), the
FKP parametrization describes rather well the leading-twist, 
leading-order anomalous part of $F_2(x,Q^2)$ at large $x$ in 
the (standard) DIS scheme. However, a number of problems remain:
\begin{Itemize}
\item The unmodified FKP expression (\ref{F2FKP}) includes extra terms 
that have no correspondence in perturbative QCD: they are neither 
leading-twist 
NLO terms nor higher-twist contributions. These terms depend on 
an additional, unphysical parameter, the light-quark masses. 
\item The FKP result is obtained in the valence approximation, 
and thus does not reproduce the correct small-$x$ behaviour of $F_2$ 
(which, actually, matters already for $x < 0.5$). 
\item A hadronic part of $F_2^\gamma$ matched to the 
FKP parametrization is not available. The recipe 
``TPC/$2\gamma$-plus-FKP" is inconsistent: since the FKP 
parametrization needs a $p_{T0} \leq 0.5\,$GeV to 
describe the high-$Q^2$ data, FKP is already large 
at $Q^2=0.71\,$GeV$^2$, where the TPC parametrization alone 
is supposed to describe the data. 
\item Since the FKP parametrization in eq.~(\ref{fqFKP})
mixes kinematic terms  between the evolved quark distribution functions 
and the direct term, it is not a parametrization of the anomalous 
quark distribution functions of the photon (at least not with 
default parameters). The quark distribution functions do  
not show the correct small-$x$ behaviour and a 
parametrization of the gluon density is missing.
\end{Itemize}

We have attempted to provide an alternative where these problems
are resolved. Our quark and gluon distribution functions obey 
the leading-logarithmic evolution equations down to very small 
$x$ ($\sim 10^{-4}$). The only free parameter besides $\Lambda$ 
is the scale $Q_0$. Moreover, for two representative values of $Q_0$ 
($0.6\,$ and $2.0\,$GeV), our anomalous PDFs are supplemented by 
parametrizations of the hadronic PDFs.  These have been obtained by 
fitting the photon structure function $F_2^\gamma(x,Q^2)$ to the 
available data.

\section{Concluding remarks}

In this paper we have shown how the PDFs of the photon can be 
decomposed into perturbative and non-perturbative components. 
Since this separation and the constraints on the 
non-perturbative part is the central result of our paper, 
we briefly repeat the main ideas of our approach and show its 
connection with an integral representation of the photon 
structure functions. Our starting point is the fact that a PDF, 
being the solution of an inhomogeneous evolution equation, can always 
be written as the sum of two terms: a particular solution 
$f_a^{\gamma,\mathrm{PT}}(x,Q^2,Q_0^2)$, which vanishes at $Q^2=Q_0^2$, 
and a general solution $f_a^{\gamma,\mathrm{NP}}(x,Q^2,Q_0^2)$
of the corresponding homogeneous evolution equation, that needs 
an input at $Q^2=Q_0^2$. 
For $Q_0$ sufficiently large, such that higher-twist ($1/Q^2$) 
effects become negligible, the anomalous distribution function 
defined in eq.~(\ref{angamsol}) is a particular solution. Hence 
we can identify $f_a^{\gamma,\mathrm{PT}}(x,Q^2,Q_0^2) 
= f_a^{\gamma,\mathrm{anom}}(x,Q^2,Q_0^2)$.  

To constrain $f_a^{\gamma,\mathrm{NP}}(x,Q^2,Q_0^2)$ we first observe 
that the anomalous distribution function of a parton $a$ within the 
photon is the convolution of two factors. The first one, 
$P^\gamma_{\qqbar}(k) = (\alphaem/2\pi) 2 e_{\q}^2 {\d} k^2/k^2$
gives the  probability for the photon to branch 
into a $\qqbar$ state at some (perturbatively high) scale $k$. 
The second factor is the distribution function of the parton $a$ 
within this $\qqbar$ state. This ``state" distribution function 
obeys the homogeneous evolution equation (with the calculable input 
(\ref{fxinit})). 

This factorization of $f_a^{\gamma,\mathrm{PT}}$ suggests a similar 
one for $f_a^{\gamma,\mathrm{NP}}(x,Q^2,Q_0^2)$. 
Below the scale $Q_0$, the $\gamma\rightarrow\qqbar$ 
transition can no longer be calculated perturbatively but may be 
approximated by fluctuations of the photon into vector mesons 
with probabilities $P^\gamma_V(k^2 = m_V^2) = 4\pi\alphaem/f_V^2$.  
Hence the analogue of the $k$-integral above $Q_0$ of the continuous 
spectrum of perturbative (``anomalous") states is the discrete sum of 
vector mesons, with the state distribution functions  
$f_a^{\gamma,\qqbar}$ replaced by the PDFs $f_a^{\gamma,V}$ of the 
$\qqbar$ ``bound states" $V$ of the photon. 
As long as the shape of these distribution functions are treated as 
free parameters (to be fixed from data), 
$f_a^{\gamma,\mathrm{NP}} = f_a^{\gamma,\mrm{VMD}}$ 
defined in eq.~(\ref{VMDgammaPDF}) is indeed a general solution of the 
homogeneous evolution equations.  

Our decomposition (\ref{VMDsum}) also follows from the representation 
of the (moments of the) 
photonic PDFs as a dispersion integral in the photon mass 
($P^2$ is the photon virtuality) \cite{Bj}
\begin{equation}
  f_a^\gamma(n,Q^2,P^2) = \int_0^{Q^2} \frac{{\d} k^2}{k^2 + P^2} 
  \rho_a(n,Q^2,k^2)
\ .
\label{dispersion}
\end{equation} 
Rather than describing the dispersion integral as the difference between  
a ``point-like" part (contribution from the upper limit) and a 
``hadronic" part (contribution from the lower limit), 
it is more natural to separate short-distance 
and long-distance parts by a scale $Q_0$ since the weight function $\rho_a$ 
possesses the scaling-violation pattern typical of ordinary hadronic 
PDFs \cite{Bj}. Our solution follows from the ansatz 
\begin{eqnarray}
 \rho_a(n,Q^2,k^2) & = & \frac{\alphaem}{\pi} e_{\q}^2 
         f_a^{\gamma,\qqbar}(n,Q^2,k^2) \Theta(k^2-Q_0^2) 
\nonumber\\ & & 
  + \sum_V \frac{4\pi\alphaem}{f_V^2} 
   \delta'\left(1 - \frac{k^2}{m_V^2}\right)
   f_a^{\gamma,V}(n,Q^2,Q_0^2)
\ .
\label{rhoansatz}
\end{eqnarray}
The scale $Q_0$ is arbitrary; hence the $Q_0$ dependence 
of the anomalous part must be cancelled by that of the hadronic one. 
For example, in the generalized vector-meson-dominance ansatz 
(\ref{Kfactor}) that we are using, the relation between the two 
parts is given by eq.~(\ref{normrelation}). 

For a given, chosen $Q_0$, 
the only unknown (non-perturbative) pieces of the photon structure 
functions and the PDFs of the photon are the valence, sea, and gluon 
input distributions and the value of $K$ in eq.~(\ref{Kfactor}). 
In order to cover the uncertainties associated with our approach 
we have presented two extreme analyses. 
In the first we restricted VMD to the well-established 
$\rho^0$, $\omega$, $\phi$ states only (i.e.\ $K=1$) and fixed 
$Q_0=0.6\,$GeV as obtained from our $\gamma\p$ analysis. Here the main 
theoretical error arises from the use of perturbation theory down to  
rather low values of $Q^2$. The spirit of the second analysis was 
opposite: take $Q_0$ well within the perturbative domain ($Q_0=2\,$GeV) 
at the expense of parametrizing the effects of additional vector mesons 
by a simple factor $K$ to be fitted to the data.

We have presented self-contained formulae and
parametrizations that allow the PDFs and $F_2$ of the photon
to be evaluated as a function of $x$ and $Q^2$, and also for
non-vanishing target virtuality $P^2$. 
A program with this information already encoded is obtainable
on request from the authors. The anomalous PDFs are 
parameter-free; more precisely, the parametrization depends 
analytically on $Q_0$, $\Lambda$, and $P^2$ (besides, of course ,
$Q^2$ and $x$). Similarly, the state distributions can be evaluated 
for arbitray $x$, $Q^2$, $k^2$, and $\Lambda$. 
Based on fits to the available $F_2^\gamma(x,Q^2)$ data, 
four different sets of hadronic PDFs are provided in order 
to illustrate the uncertainties in current PDF determinations. 
The sets differ in the value of $Q_0$ ($0.6$ and $2\,$GeV)  
and the data included (in both cases only data above $Q_0^2$ are included 
in the fit), and the factorization scheme. The latter dependence 
is formally of NLO accuracy, but is found to be numerically significant 
in LO analyses.  

We have outlined the small-$x$ behaviour of the PDFs and 
also discussed the description of the hadronic 
event properties in photon-induced reactions. A proper treatment 
requires knowledge beyond that of the usual inclusive PDF 
parametrizations. 
Moreover, we have investigated how the photon-to-$\qqbar$ 
splitting probability enters the eikonalization of jet cross sections. 
We have derived a formula that ensures that the various contributions 
are not counted twice. We have shown that it is mandatory to  
change from the inclusive PDFs to the state distributions.  

Finally we have investigated the limitations of the approximation 
of $F_2^\gamma(x,Q^2)$ by the ``FKP-plus-TPC$/2\gamma$" expression. 
We found that (with correct parameter choice) the FKP formula  
is a good parametrization of the anomalous part of $F_2^\gamma(x,Q^2)$ 
at large $x$ but fails for $x < 0.5$. Concerning the parametrization 
of the hadronic part, we demonstrated the effects of the necessary 
$Q^2$ evolution of the TPC/$2\gamma$ parametrization and, 
most importantly, pointed out the strong correlation between the 
cutoff scale $p_{T0}$ of the FKP formula and the (size and shape of the) 
hadronic part: if $p_{T0}$ is varied only in the anomalous part, it 
introduces an artificial dependence on an unphysical parameter, which 
must not be there.

\subsection*{Acknowledgement}
We thank A.\ Vogt for providing us with tables of $F_2^{\gamma}$ data. 

\begin{thebibliographys}{99}
 
\bibitem{Berger87}
C.\ Berger and W.\ Wagner, Phys.\ Rep.\ {\bf 146} (1987) 1

\bibitem{Kolanoski88}
H.\ Kolanoski and P.\ Zerwas, 
in {\it High-energy electron--positron physics} (World Scientific, 
Singapore, 1988), eds.\ A.\ Ali and P.\ S\"{o}ding, p.\ 695

\bibitem{Vieira91}
J.H.\ Da Luz Vieira and J.K.\ Storrow, 
Z.\ Phys.\ {\bf C51} (1991) 241

\bibitem{Borzumati93}
F.M.\ Borzumati and G.A.\ Schuler, Z.\ Phys.\ 
{\bf C58} (1993) 139

\bibitem{Vogt94}
A.\ Vogt, in {\it Proc.\ of the Workshop on 
Two-Photon Physics at LEP and HERA} 
(Lund, Sweden, 1994), eds. G.\ Jarlskog and L.\ J\"{o}nsson, p. 141

\bibitem{Brodsky94}
S.J.\ Brodsky and P.M.\ Zerwas, SLAC-PUB-6571, July 1994; to appear in 
{\it Proc.\ of the Workshop on $\gamma\gamma$ Colliders}, 
LBL, USA, March 1994

\bibitem{Miller94}
D.J.\ Miller, in {\it Proc.\ of the Workshop on 
Two-Photon Physics at LEP and HERA} 
(Lund, Sweden, 1994), eds. G.\ Jarlskog and L.\ J\"{o}nsson, p. 4

\bibitem{LAC} 
M.\ Drees and K.\ Grassie,  Z.\ Phys.\ {\bf C28} (1985) 451;
\hfill\\
H.\ Abramowicz, K.\ Charchula and A.\ Levy, 
Phys.\ Lett.\ {\bf B269} (1991) 458;
\hfill\\
K.\ Hagiwara, M.\ Tanaka, I.\ Watanabe and T.\ Izubuchi, 
 ``Gluon and charm distributions in the photon",
 KEK preprint 93-160, March 1994 (hep-ph 9406252)

\bibitem{GRV}
M.\ Gl\"{u}ck, E.\ Reya and A.\ Vogt, Phys.\ Rev.\ 
{\bf D46} (1992) 1973

\bibitem{AFG}
P.\ Aurenche, M.\ Fontannaz and J.-Ph.\ Guillet, 
LAPP preprint ENSLAPP-A-435-93-REV, 1993

\bibitem{GS}
L.E.\ Gordon and J.K.\ Storrow, Z.\ Phys.\ {\bf C56} 
(1992) 307

\bibitem{Schuler93}
G.A. Schuler and T. Sj\"ostrand, Phys. Lett. {\bf B300} 
(1993) 169 and 
Nucl. Phys. {\bf B407} (1993) 539

\bibitem{Walsh}
T.F.\ Walsh and P.M.\ Zerwas, Nucl.\ Phys.\ {\bf B41} (1972) 551 and 
Phys.\ Lett.\ {\bf B44} (1973) 195;
\hfill\\
R.\ Kingsley, Nucl.\ Phys.\ {\bf B60} (1973) 45 4

\bibitem{Uematsu}
T.\ Uematsu and T.F.\ Walsh, Phys.\ Lett.\ {\bf B101} (1981) 263 
and Nucl.\ Phys.\ {\bf B199} (1982) 93;
\hfill\\
G.\ Rossi, Univ.\ California at San Diego preprint UCSD-10P10-227
(1983) and Phys.\ Rev.\ {\bf D29} (1984) 852

\bibitem{Kim93}
S.M.\ Kim and T.F.\ Walsh, Univ.\ of Minnesota preprint 
UMN-TH-1111-92, September 1992, hep-ph-9308267

\bibitem{Drees94} 
M.\ Drees and R.M.\ Godbole, Madison preprint MAD/PH/819, 
March 1994

\bibitem{Aurenche94}
P.\ Aurenche, J.-Ph.\ Guillet, M.\ Fontannaz, Y.\ Shimizu, 
J.\ Fujimoto and K.\ Kato, in {\it Proc.\ of the Workshop on 
Two-Photon Physics at LEP and HERA} 
(Lund, Sweden, 1994), eds. G.\ Jarlskog and L.\ J\"{o}nsson, p. 267

\bibitem{Gluck94}
M.\ Gl\"{u}ck, E.\ Reya and M.\ Stratmann, Dortmund preprint 
DO-TH 94/14, August 1994

\bibitem{TPC}
TPC/$2\gamma$ collab., H.\ Aihara et al., Z.\ Phys.\ {\bf C34} (1987) 1  
 and Phys.\ Rev.\ Lett.\ {\bf 58} (1997) 97

\bibitem{FKP}
J.H.\ Field, F.\ Kapusta and L.\ Poggioli, Phys.\ Lett.\ 
{\bf B181} (1986) 362 and Z.\ Phys.\ {\bf C36} (1987) 121;
\hfill\\
F.\ Kapusta, Z.\ Phys.\ {\bf C42} (1989) 225

\bibitem{PWZ83}
C.\ Peterson, T.F.\ Walsh and 
P.M.\ Zerwas, Nucl.\ Phys.\ {\bf B229} (1983) 301

\bibitem{Odorico}
R.\ Odorico, Phys. Lett. {\bf B102} (1981) 341 and 
Comput.\ Phys.\ Commun.\ {\bf 25} (1982) 253

\bibitem{James}
F.\ James, ``MINUIT --- Function Minimization and Error Analysis", 
version 92.1 (March 1992), CERN Program Library D506, CERN, Geneva 1992

\bibitem{Yuri}
Yu.L.\ Dokshitzer and D.V.\ Shirkov, LU TP 93-19 

\bibitem{Budnev}
V.M.\ Budnev et al., Phys.\ Rep.\ {\bf 15} (1975) 181

\bibitem{Hill}
C.T. Hill and G.G. Ross, Nucl. Phys. {\bf B148} (1979) 373

\bibitem{Bj}
 J.D.\ Bjorken, ``Two topics in quantum chromodynamics", 
 SLAC-PUB-5103, December 1989

\bibitem{JADE}
JADE collab., W.\ Bartel et al., Z.\ Phys.\ {\bf C24} (1984) 231

\bibitem{PLUTO}
PLUTO collab., Ch.\ Berger et al., Z.\ Phys.\ {\bf C26} (1984) 353 and  
 Nucl.\ Phys.\ {\bf B281} (1987) 365

\bibitem{TASSO}
TASSO collab., H.\ Althoff et al., Z.\ Phys.\ {\bf C31} (1986) 527

\bibitem{AMY}
AMY collab., T.\ Sasaki et al., Phys.\ Lett.\ {\bf 252} (1990) 491

\bibitem{CELLO}
CELLO collab., H.J.\ Behrend et al., contributed paper to the XXVth 
  Int.\ Conf.\ on HEP, Singapore, 1990

\bibitem{TOPAZ}
TOPAZ collab., K.\ Muramatsu et al., Phys.\ Lett.\ {\bf B332} (1994) 477

\bibitem{OPAL}
OPAL collab., R.\ Akers et al., Z.\ Phys.\ {\bf C61} (1994) 199
 
\end{thebibliographys}

\clearpage

\end{document}